\documentclass[aoas]{imsart}

\RequirePackage{amsthm,amsmath,amsfonts,amssymb}
\RequirePackage[authoryear]{natbib}
\RequirePackage[colorlinks,citecolor=blue,urlcolor=blue]{hyperref}
\RequirePackage{graphicx}
\RequirePackage{subfig}
\RequirePackage{multirow}
\RequirePackage{float}
\RequirePackage{url}

\startlocaldefs
\theoremstyle{plain}

\theoremstyle{definition}

\endlocaldefs

\begin{document}

\begin{frontmatter}
\title{Dynamic Prediction of the Target Survival Time in Metastatic Solid Tumor Cancer Clinical Trials}
\runtitle{Dynamic Prediction of Target Survival Time}

\begin{aug}
\author[A]{\fnms{Sidi}~\snm{Wang}\ead[label=e1]{sidiwang@umich.edu}\orcid{0000-0003-4838-0842}},
\author[A]{\fnms{Kelley}~\snm{Kidwell}\ead[label=e2]{kidwell@umich.edu}\orcid{0000-0002-1717-4483}},
\author[B]{\fnms{Bo}~\snm{Huang}\ead[label=e3]{Bo.Huang@pfizer.com}\orcid{0000-0002-3088-9328}}
\and
\author[B]{\fnms{Satrajit}~\snm{Roychoudhury}\ead[label=e4]{Satrajit.Roychoudhury@pfizer.com}\orcid{0000-0003-4001-3036}}
\address[A]{University of Michigan, Department of Biostatistics \textbf{}\printead[presep={ ,\ }]{e1,e2}}

\address[B]{Pfizer Inc.\textbf{}\printead[presep={,\ }]{e3,e4}}
\end{aug}

\begin{abstract}
Overall survival (OS) is the gold standard for assessing patient benefit and cost-effectiveness of new cancer drugs. However, it is often difficult to use OS as the primary endpoint in randomized clinical trials (RCTs) for patients with metastatic cancer due to multiple reasons. In recent years, progression-free survival (PFS) has increasingly been used as the primary endpoint in metastatic cancer RCTs to accelerate development. However, regulatory authorities often seek mature OS data for approval. Therefore, it is critical to determine the target time when OS data are expected to be mature for reliable statistical inference. Motivated by an advanced renal cell carcinoma (RCC) clinical trial, we develop and investigate different prediction models leveraging information from disease progression to improve target OS prediction times. We propose a multivariate joint modeling approach considering components of progression and OS and extend three models commonly used for association to be used for OS prediction. To the best of our knowledge, this is the first comprehensive statistical study exploring the prediction of OS using different levels of information on disease progression and illustrating these models using a real, complex dataset. Our findings have significant implications for OS prediction.
\end{abstract}


\end{frontmatter}

\section{Introduction}

In recent years, many cancer drugs have received regulatory approvals based on results from phase 3 randomized clinical trials (RCTs) with progression-free survival (PFS) as the primary endpoint. Though PFS is well accepted as an intermediate endpoint for many cancer types, improvement in overall survival (OS) remains the clinical gold standard for assessing patient benefit \citep{tang2007surrogate, driscoll2009overall, methy2010surrogate, grigore2020surrogate} and cost-effectiveness \citep{royle2023overall} of a new drug. Planning a statistically well-powered OS analysis in practice is often challenging due to patients switching to alternative treatments after progression, non-response to treatment, starting other anti-cancer therapies, or being lost to follow-up. The timing of the RCT primary analysis with PFS as the endpoint is primarily driven by the number of patients who progressed. Therefore, at the time of the primary analysis, OS data are often not mature enough for meaningful statistical inference regarding treatment effects due to a low number of deaths. If PFS is statistically significant in the final analysis, regulatory agencies often request one or more updated OS analyses once the survival data are more mature. Thus, proper planning is necessary to understand when mature OS data will be available.
 
Model-based prediction of OS time for trial participants helps research teams allocate resources efficiently, plan future OS analyses accurately, and assess the likelihood of demonstrating the survival benefit of an experimental drug in the analysis. Accurate OS predictions not only facilitate the effective use of limited healthcare resources but also may benefit the patient in allowing the patient, their clinician, and family to make suitable plans for the remainder of the patient's life \citep{mackillop1997measuring}. \cite{sborov2019impact} demonstrated that when oncologists inaccurately predict OS, patients with advanced cancer are more likely to receive aggressive end-of-life care, which often contradicts the patients' wishes. 

Beyond drug approval and patient benefit, accurate OS predictions play a key role in determining the cost-effectiveness of a new drug and whether it is recommended for use in standard of care and reimbursement. Cost-effectiveness evaluations of new cancer treatments in health technology assessments rely on model-based extrapolation \citep{dias2011nice, latimer2011nice}. Various national health authorities, including the National Institute for Health and Care Excellence (NICE), the Pharmaceutical Benefits Advisory Committee, and the Canadian Agency for Drugs and Technologies in Health, provide guidance on appropriate extrapolation techniques for individual trials using parametric and semi-parametric distributions \citep{canadian2006guidelines, latimer2011nice, pharmaceutical2016guidelines}. \cite{heeg2022novel} have discussed a pragmatic two-step approach to select the optimal model and its corresponding distribution to ensure that the most plausible settings are being used for health technology assessment submissions. \cite{henderson2001accuracy} further provide examples of how accurate survival predictions have financial implications for insurance programs and health authorities. However, none of these approaches explored the relationship between progression and death to improve prediction accuracy.

When PFS is used as the primary endpoint, it captures the process of progression across target, non-target, and new lesions and death. That is, PFS is a composite endpoint defined as the time from randomization until progressive disease (PD) or death from any cause. In solid tumor studies, PD is evaluated using the Response Evaluation Criteria in Solid Tumors \citep{eisenhauer2009new}, commonly referred to as RECIST. The following four components comprise PFS: 1) measurement of the target lesion, which is captured as longitudinal continuous data; 2) time to worsening of the non-target lesion; 3) time to the emergence of a new lesion; and 4) time to death from any cause. Depending on the cancer type, each of these components has a different predictive degree of OS. For example, \cite{stein2013survival} showed that in Metastatic Renal Cell Carcinoma (RCC) the worsening of a non-target lesion or the appearance of a new lesion are more predictive of OS benefit than the sum of the longest tumor diameter of the target lesion. The current practice of OS prediction often ignores the disease progression process and models only the number of deaths.

In this manuscript, we develop and evaluate several model-based approaches to forecast the survival (death) times of trial participants ``still alive" leveraging information about disease progression and important baseline factors. Quantifying the relationship between disease progression and survival has garnered interest in both statistical and clinical literature over time. Among some of the notable ones, \cite{fleischer2009statistical} employed exponential time-to-event distributions to describe the dependency structure between OS and PFS. This is further expanded by \cite{fu2013joint, weber2019quantifying} and \cite{meller2019joint} by utilizing copula and multi-state models to jointly analyze PFS and OS without a strict parametric assumption for the marginal survival distributions of PFS and OS. \cite{shukuya2016relationship} investigated the correlation between median PFS and median OS, concluding that both tumor response and PFS are significant predictors of OS. Other methodologies for OS prediction include joint modeling of the longitudinal tumor size data of target tumor and survival data (\cite{claret2009model, claret2013evaluation, wang2009elucidation, bruno2014evaluation, zecchin2016models}, and \cite{lim2019predicting}). Meanwhile, \cite{yu2020new} jointly modeled the dynamics of target lesions and the progression of non-target lesions to predict PFS. However, most of the literature in this area is primarily focused on improving the estimation of OS rather than predicting the future survival time of patients who are ongoing in the trial. 

To our current knowledge, there is no existing method that formally combines all three components of disease progression along with additional uncertainties associated with death while predicting survival. We propose a multivariate joint modeling approach in this paper to fill that gap. Further, popular methods to analyze PFS and OS jointly, like the shared parameter joint model, copula and multi-state model methods, have not previously been developed and illustrated for OS prediction. Hence, we extend the shared parameter joint model, copula and multi-state models used primarily for association to be used for OS prediction. We compare OS predictions across these three novel methods and to a traditional marginal Weibull baseline hazard model of OS.

\section{Motivating Example: Phase 3 Study in Renal Cell Carcinoma} \label{sec:example}
\begin{figure}
    \centering
    \includegraphics[width=12cm]{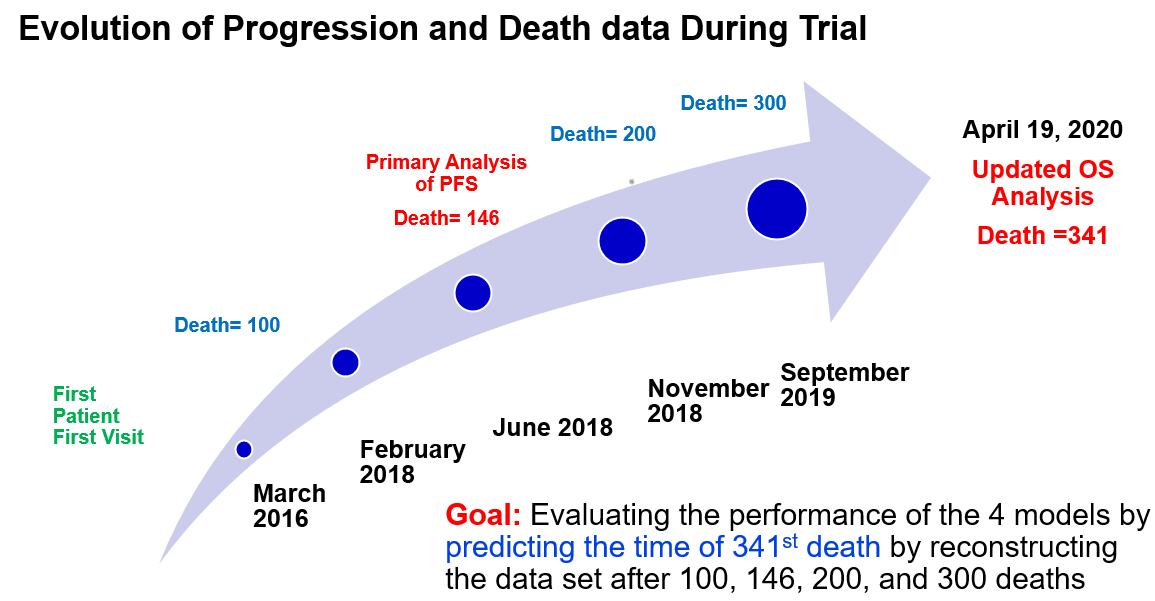}
    \caption{Evolution of progression and death data during trial\label{fig:evolution}}
\end{figure}

The methods outlined in this paper are inspired by a phase 3, randomized, open-label, parallel-arm clinical trial \citep{motzer2019avelumab}. Treatment-naive adult participants with advanced RCC were randomized (1:1) to receive either the experimental drug or the active comparator, which was the standard of care. A key inclusion criterion was the presence of at least one measurable lesion, as defined by RECIST version 1.1. Tumor assessments were performed using computed tomography or magnetic resonance imaging at baseline, every six weeks post-randomization for the first 18 months, and subsequently every 12 weeks until confirmed disease progression. Important baseline demographic and disease characteristics were evenly distributed between the two treatment groups. The study had two primary endpoints: PFS and OS. The primary analysis was scheduled when at least 397 PFS events occurred (Figure \ref{fig:evolution}). At the time of the primary PFS analysis, only 146 deaths were observed, which was immature to draw any reasonable conclusion about the OS benefit provided by the new drug. Therefore, an updated analysis of OS was planned after 341 deaths in the trial. The trial findings indicated that patients administered the experimental drug experienced a markedly extended PFS compared to those given the active comparator. It is crucial to note that while we have utilized the data structure from a real-life trial, the actual values presented here are simulated based on published summary statistics, both for proprietary considerations and to ensure proper protection of patient data.

We aim to use the observed data to construct a predictive tool that reliably estimates the time of the $n$th death in the trial. In addition to timing, we are also interested in calculating the likelihood of demonstrating survival benefits in the updated OS analysis. This information is strategically and ethically important for retaining patients on the experimental drug. Specifically, we employ data on participants' baseline characteristics, treatment group, current survival status, and the details of tumor progression, i.e., target lesion measurements and the progression of disease of non-target and new lesions. We develop models in the setting of RCC since PFS is known to be a good predictor of OS \citep{heng2011progression}, but the proposed model could be applied to other similar metastatic cancer settings.

Supplementary Figure 1 shows significant variability between individuals for the sum of the longest diameters of target lesions. Therefore, it is difficult to draw any conclusions about treatment without systematically modeling the variability. The Kaplan-Meier plots of non-target lesions and new lesions are shown in Supplementary Figure 2b and 2c. We observe a difference between the two treatment arms from both plots: patients who received the experimental drug had a longer time to PD considering non-target lesion and new lesion assessments. A similar trend is also observed in the Kaplan-Meier plots of OS (Supplementary Figure 2d) and PFS (Supplementary Figure 2e). 

\section{Statistical Models for Milestone Prediction of Survival Endpoints}
\label{sec:method}

We present four different methods (Sections \ref{sec:jm}, \ref{sec:spjm}, \ref{sec:method_copula}, and \ref{sec:multistate}) for predicting milestone survival times based on disease progression information. These models use the components of tumor progression in different formats (e.g., as each component separately or as a composite endpoint) along with the available death data to predict the time of death for ongoing patients. To understand the efficiency gain from progression data, we have included a parametric model (Section \ref{sec:method_standard}) that uses death information marginally, ignoring progression status. 

\subsection{Prediction Model Based on Jointly Modeling Different Components of Disease Progression}
\label{sec:jm}

We start with a new multivariate joint modeling approach (baveJM) that models the association between each component of disease progression and survival. Three separate models considering the association between survival and i) the sum of the longest diameter of target lesions, ii) the time to non-target lesion PD, and iii) the time to new lesion PD are proposed to explain the underlying heterogeneity of tumor progression dynamics and prevent information loss. In addition, we include a marginal model of current OS to take into account other intercurrent events (e.g., cross-over of control patient to treatment arm after progression) that can potentially affect the long-term survival of patients. 

\begin{figure}
\centering
\includegraphics[width=14cm]{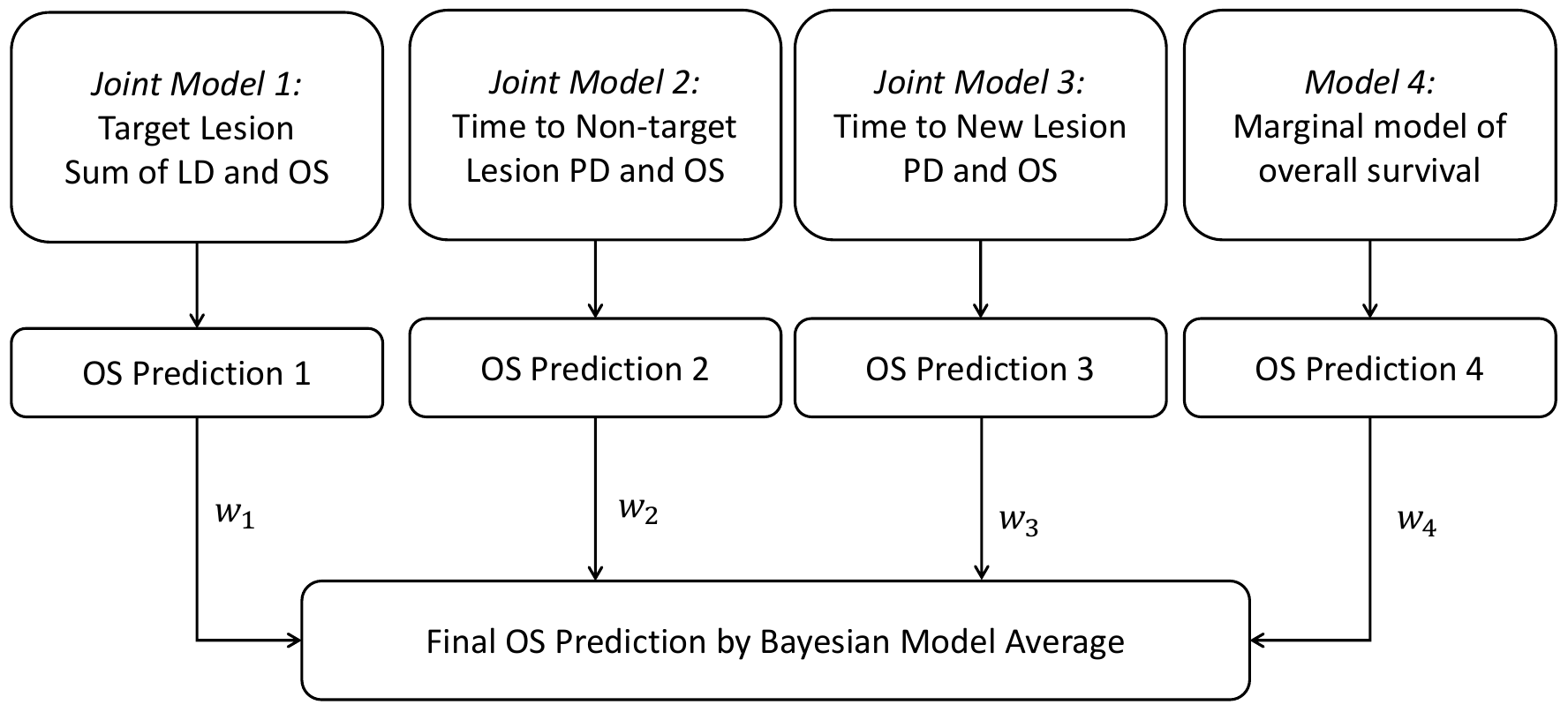}
\caption{Multivariate joint modeling structure.} 
\label{fig:JM}
\end{figure}

The proposed baveJM provides real-time OS predictions from each trial participant ``still alive" in two steps. First, we derive four sets of intermediate predictions based on the four models by fully encapsulating the association between progression and death. In the second step, we consolidate the OS predictions from all four models simultaneously using Bayesian model averaging (BMA) \citep{hoeting1999bayesian}. The structure of baveJM is illustrated in Figure \ref{fig:JM}. 

\subsubsection{Joint Model 1: the Association between Target Lesion and OS} \label{sec:TL}

A shared parameter model is used to explore the association between target lesion measurements and OS. We consider a set of $n$ subjects followed over an interval of time [0,$\tau$]. Let $z_{ij}$ be the target lesion measurement (or \% change from baseline in the sum of the longest diameter of target lesions) at time $t_{ij}$ (for the $i$-th participant at the $j$-th visit) where $i=1,...,n$ and $j=1,...,k_i$. We assume the target lesion measurement time $t_{ij}$ is non-informative so that it is independent of the longitudinal measurement and event-time processes for OS. This is a reasonable assumption as the tumor assessment timings were often specified in the protocol before study starts. We define a latent zero-mean Gaussian process $W_i(t)$ for the $i$-th participant, independent of different participants. The following sub-models link the joint model of lesion measurements and OS:
\begin{eqnarray*}
Y_{i}(t) \sim N(\mu_{i}(t), \sigma^{2}_{\epsilon}),
\end{eqnarray*}
where $Y_{i}(t)$ and $\sigma_{\epsilon}$ are the target lesion measurement for subject $i$ at time $t$ and variability associated with it. Moreover, the lesion measurement process $\mu(t)$ is modeled by the linear model
\begin{eqnarray*}
\mu_{i}(t) &=& \textbf{X}_i(t)' \boldsymbol{\beta}_{\mu}+W_{i}(t), \\
W_{i}(t) &=& b_{1i}+b_{2i}t.
\end{eqnarray*}
Here, $\textbf{X}_i(t)$ refers to a time-dependent covariate matrix for subject $i$'s lesion measurement. Finally, $W_i(t)$ is a Gaussian process with ($b_{1i},b_{2i}$) being zero-mean bivariate Gaussian variables with variances $\sigma_1^2$ and $\sigma_2^2$ respectively, and correlation coefficient $\rho$. 

The OS process at time $t$ is modeled by a Weibull model:
\begin{eqnarray*}
\lambda(t) &=& \lambda_0(t) \exp(\textbf{Z}_i' \boldsymbol{\beta}_{\rm OS}+\lambda \mu_{i}(t))\\
 &=& \alpha_{\rm OS,TL} \gamma_{\rm OS} t^{\alpha_{\rm OS,TL}-1} \exp(\textbf{Z}_i' \boldsymbol{\beta}_{\rm OS}+\lambda \mu_{i}(t)),
\end{eqnarray*}
where $\textbf{Z}_i$ is a covariate matrix for OS, $\gamma_{\rm OS}$ and $\alpha_{\rm OS,TL}$ are the scale and shape parameters
of the Weibull distribution. 

\subsubsection{Joint Models 2 and 3: the Association Between OS and Time to Worsening of Non-Target Lesion or Time to Appearance of New Lesion} \label{sec:NT}

Based on RECIST 1.1, the detailed tumor measurements of non-target and new lesions are not available at each visit. For example, we only know the current status of the non-target lesion (stable/worsen/disappear) and the appearance of a new lesion (yes/no). Moreover, this information is only available until the date of progression. Therefore, we use the time to worsening of non-target lesion (or time to appearance of new lesion) for the prediction model. The association between time to worsening of non-target lesion or time to appearance of new lesion and OS are explored using separate bivariate distributions. We used a copula-based model to capture the dependence between these survival times. Copulas are commonly used for modeling the dependence between random variables. They are continuous multivariate cumulative distributions with each random variable following a uniform marginal distribution on the interval $[0, 1]$. Sklar's theorem states that any multivariate joint distribution can be written using univariate marginal distribution functions and a copula describing the variables' dependence structure \citep{sklar1959fonctions}. 

We propose a Clayton copula function  \citep{clayton1978model} to model dependencies between time to worsening of non-target lesion (or time to appearance of the new lesion) and OS. The proposed class of Archimedean copula can capture strong dependence in the left tail. The relationship between time to worsening of non-target lesion (NT) (or time to appearance of new lesion (NL)) and OS is captured by a single parameter $\eta_{NT}$ (or $\eta_{NL}$), a parameter that can be directly linked to Kendall's tau, effectively characterizing their dependence. In the subsequent paragraphs, we formally define the copula model between time to worsening of non-target lesion (or time to appearance of new lesion) and OS. The model can be specified as
\begin{eqnarray*}
S_1(t_{\rm NT},
t_{\rm OS}|\textbf{Z}) &=& \{S_{\rm NT}(t_{\rm NT}|\textbf{Z})^{-\eta_{\rm NT}}
+S_{\rm OS}(t_{\rm OS}|\textbf{Z})^{-\eta_{\rm NT}}-1\}^{-1/\eta_{\rm NT}},\\
S_2(t_{\rm NL},
t_{\rm OS}|\textbf{Z}) &=& \{S_{\rm NL}(t_{\rm NL}|\textbf{Z})^{-\eta_{\rm NL}}
+S_{\rm OS}(t_{\rm OS}|\textbf{Z})^{-\eta_{\rm NL}}-1\}^{-1/\eta_{\rm NL}}. 
\end{eqnarray*}
Here $t_{NT}$, $t_{NL}$, and $t_{OS}$ denote times to worsening of non-target lesion, time to appearance of new lesion, and time to death (OS), respectively. $\textbf{Z}$ denotes the set of potential disease modifiers or baseline covariates for the prediction model. $\eta_{\rm NT}, \eta_{\rm NL}>0$ measures the correlation between time to worsening of non-target lesion or time to appearance of new lesion and OS. A large value of $\eta_{\rm NT}$ or $\eta_{\rm NL}$ represents a high correlation. When $\eta_{\rm NT}$ or $\eta_{\rm NL}$ goes to 0, the correlation approaches 0, and when $\eta_{\rm NT}$ or $\eta_{\rm NL}$ goes to $\infty$, the correlation converges to 1. Moreover, marginal survival probabilities are calculated using proportional hazard models expressed as:
\begin{eqnarray*}
S_{\rm A}(t_{\rm iA}|\mathbf{Z_{i}}) &=& \mathrm{exp}(-\int_{0}^{t} \lambda_{\rm A}(t_{iA}|\mathbf{Z_{i}})), \;\; \mbox{A= NT, NL, OS; $i=1,...,n$}, \\
\lambda_{\rm A}(t_{iA}|\mathbf{Z_{i}}) &=& \lambda_{0, \rm A}(t_{iA})\mathrm{exp}(\mathbf{Z_{i}}' \boldsymbol{\beta_{\rm A}}),\\
\lambda_{0, \rm A}(t_{iA}) &=& \gamma_{\rm iA} t_{A}^{\alpha_{\rm A}}
\mathrm{exp}(\mathbf{Z_{i}}' \boldsymbol{\beta_{\rm A}} ), \\
\gamma_{\rm iA}&=&\gamma_{A} + b_{iA}.
\end{eqnarray*}
Here, we introduce random effects for the scale parameter of the Weibull model ($b_{iA}$) to take into account the additional heterogeneity in the patient population.

Note that the censoring mechanism is associated with the time to worsening of non-target lesions, time to appearance of new lesions, and OS. The details of the likelihood construction under each joint model and implementation are provided in supplemental materials. 

\subsubsection{Model 4: Marginal Model for OS}\label{sec:marginal}

Finally, we introduce a marginal OS model to account for the risk of death not explained by disease progression. This includes scenarios when patients die without any evidence of disease progression or a change in risk of death due to exposure to other anti-cancer treatments after progression. A Weibull regression model similar to the marginal OS models mentioned in previous sections is used in this context.

\subsubsection{Prediction: Posterior Predictive Distribution (PPD)} \label{sec:PPD}
For each of the models in Sections \ref{sec:TL} - \ref{sec:marginal}, we utilize the PPD technique \citep{gelman2014bayesian} to generate OS predictions. The PPD represents the distribution of potential unobserved values based on the observed values, and it follows this structure:
$$p(y_{pred}|y) = \int{p(y_{pred},\theta|y)d\theta} = \int{p(y_{pred}|\theta,y)p(\theta|y)d\theta} = \int{p(y_{pred}|\theta)p(\theta|y)d\theta}.$$
This formulation leverages the conditional independence between $y$ (observed data), $y_{pred}$ (unobserved data), and $\theta$ (parameters). The PPD can be understood as an average of conditional predictions over the posterior distribution of $\theta$. During the MCMC process, for each sampled $\theta$ from the posterior distribution, a corresponding $y_{pred}$ sample is obtained. 

To produce PPD samples for OS using the models presented in the previous sections, we used \texttt{JAGS} software \citep{plummer2017jags} along with the \texttt{rjags} package for implementation in the R environment.

\subsubsection{Bayesian Model Averaging (BMA)}
We propose a Bayesian Model Averaging (BMA) method to combine OS predictions from the four models described in Sections \ref{sec:TL} - \ref{sec:marginal}. In prognostic modeling, it is standard to generate predictions by selecting a single model based on certain metrics. In the context of oncology, different PFS components may be more correlated with OS under various tumor types or patient groups, thus offering more accurate OS predictions. Here, we use BMA to calculate the weighted average of the predicted OS under the three joint models and one marginal model for the OS defined above. The weights indicate the plausibility of each model being the most predictive model. We denote the models defined in Sections \ref{sec:TL}, \ref{sec:NT}, and \ref{sec:marginal} by $M_T$, $M_{NT}$, $M_{NL}$, and $M_{OS}$, respectively. Then by BMA, the final predicted probability of OS follows
\begin{eqnarray*}
P(Y_{pred}|Data)&=&P(Y_{pred.t}|M_T,Data)P(M_T|Data)\\
&+&P(Y_{pred.nt}|M_{NT},Data)P(M_{NT}|Data)\\
&+&P(Y_{pred.nl}|M_{NL},Data)P(M_{NL}|Data)\\
&+&P(Y_{pred.os}|M_{OS},Data)P(M_{OS}|Data),
\end{eqnarray*}
where $P(M_T|Data)$, $P(M_{NT}|Data)$, $P(M_{NL}|Data)$, and $P(M_{OS}|Data)$ are model weights. In general, if we have a total of $Q$ models, the model weight for model $q$ at the $t$th MCMC iteration is  
$$w^{(t)}_q=P(M_q|Data,\theta^{(t)})=\frac{P(M_q,Data,\theta^{(t)})}{P(Data,\theta^{(t)})} =\frac{P(Data|M_q,\theta^{(t)})P(\theta^{(t)}|M_q)P(M_q)}{P(Data,\theta^{(t)})},$$
where $\theta^{(t)} = (\theta_1^{(t)}, \theta_2^{(t)},...,\theta_Q^{(t)})$. If we set $P(\theta_h|M_h, Data)=g_h$, and $q \ne h$, then according to \cite{congdon2007model},
$$P(\theta^{(t)}|M_q)=P(\theta_q^{(t)}|M_q) \prod_{h=1,h\ne q}^Q P(\theta_h^{(t)}|M_q)=P(\theta_q^{(t)}|M_q) \prod_{h=1,h\ne q}^Q g_h.$$
Therefore, 
\begin{eqnarray*}
w^{(t)}_q&=&P(M_q|Data,\theta^{(t)})\\
&=&\frac{P(Data|M_q,\theta^{(t)}_q) P(\theta^{(t)}_q|M_q) \prod_{h=1,h\ne q}^Q g_h P(M_q)}{\sum_{k=1}^Q P(M_k,Data, \theta^{(t)})} \\
&=&\frac{P(Data|M_q,\theta^{(t)}) P(\theta_q^{(t)}|M_q) \prod_{h=1,h\ne q}^Q g_h P(M_q) }{\sum_{k=1}^Q P(Data|M_k,\theta^{(t)}) P(\theta_k^{(t)}|M_k) \prod_{h=1,h\ne k}^K g_h P(M_k)}.    
\end{eqnarray*}
Through the above formulae, we implement the BMA approach and calculate the final predicted OS. 

\subsection{Prediction Models Based on Using Shared Parameter to jointly model all Components of Disease Progression}\label{sec:spjm}

We implement the shared parameter random effect joint model (SPJM) model in a single Bayesian hierarchy \citep{vonesh2006shared, lawrence2015joint, ratcliffe2004joint}, linking the longitudinal target lesion measurements, the two progression components (time to worsening of non-target lesion and time to appearance of new lesion), and OS through shared subject–level random effects. 
For participant $i=1,\ldots,n$ observed at visits $j=1,\ldots,k_i$ at times $t_{ij}$, the target lesion measurement $\tilde{Y}_{ij}$ is modeled as
\[
\tilde{Y}_{ij}\mid \tilde{b}_{1i},\tilde{b}_{2i},\tilde{\boldsymbol{\beta}}_{0},\tilde{\sigma}_Y^2 
~\sim~ N\!\big(\tilde{\mu}_{ij},\,\tilde{\sigma}_Y^2\big),\qquad
\tilde{\mu}_{ij} ~=~ \tilde{\mathbf{X}}_i^{\top}\tilde{\boldsymbol{\beta}}_{0}\;+\;t_{ij}\big(\tilde{\beta}_{0,t}+\tilde{b}_{2i}\big)\;+\;\tilde{b}_{1i},
\]
where $\tilde{\mathbf{X}}_i$ collects covariates for the lesion process, $\tilde{\boldsymbol{\beta}}_{0}$ are fixed effects, and $\tilde{\beta}_{0,t}$ is a fixed time slope. We introduce four subject–specific latent terms: a correlated intercept–slope pair $(\tilde{b}_{1i},\tilde{b}_{2i})$ with correlation $\tilde{\rho}$, and two additional Gaussian frailties $\tilde{b}_{3i}$ and $\tilde{b}_{4i}$,
\[
\begin{pmatrix} \tilde{b}_{1i}\\[2pt] \tilde{b}_{2i} \end{pmatrix}
\sim N\!\left(
\begin{pmatrix}0\\[1pt]0\end{pmatrix},
\begin{pmatrix}
\tilde{\sigma}_{b1}^2 & \tilde{\rho}\,\tilde{\sigma}_{b1}\tilde{\sigma}_{b2}\\
\tilde{\rho}\,\tilde{\sigma}_{b1}\tilde{\sigma}_{b2} & \tilde{\sigma}_{b2}^2
\end{pmatrix}
\right),\quad
\tilde{b}_{3i}\sim N\!\big(0,\tilde{\sigma}_{b3}^2\big),\quad
\tilde{b}_{4i}\sim N\!\big(0,\tilde{\sigma}_{b4}^2\big),
\]
with $\tilde{b}_{3i}\perp(\tilde{b}_{1i},\tilde{b}_{2i})$ and $\tilde{b}_{4i}\perp(\tilde{b}_{1i},\tilde{b}_{2i},\tilde{b}_{3i})$. The two progression components are fit with Weibull regressions, with the scale modeled log–linearly using the shared effects. For the time to worsening of non–target lesion (NT), $\tilde{T}_{{\rm NT},i}$,
\[
\tilde{T}_{{\rm NT},i}\mid \tilde{\alpha}_{\rm NT},\tilde{\theta}_{{\rm NT},i}
~\sim~{\rm Weibull}\!\big(\tilde{\alpha}_{\rm NT},\tilde{\theta}_{{\rm NT},i}\big),
\]
\[
\log \tilde{\theta}_{{\rm NT},i}
~=~ \tilde{\mathbf{Z}}_{{\rm NT},i}^{\top}\tilde{\boldsymbol{\beta}}_{\rm NT}
\;+\;\tilde{\phi}_1 \tilde{b}_{1i}\;+\;\tilde{\phi}_2 \tilde{b}_{2i}\;+\;\tilde{b}_{3i},
\]
so that $\tilde{h}_{{\rm NT},i}(t)=\tilde{\alpha}_{\rm NT}\,\tilde{\theta}_{{\rm NT},i}\,t^{\tilde{\alpha}_{\rm NT}-1}$. Here $\tilde{b}{1i}$ and $\tilde{b}{2i}$ are incorporated via association parameters $(\tilde{\phi}_1,\tilde{\phi}_2)$, while $\tilde{b}_{3i}$ acts as an NT–specific frailty. For the time to appearance of new lesion (NL), $\tilde{T}_{{\rm NL},i}$,
\[
\tilde{T}_{{\rm NL},i}\mid \tilde{\alpha}_{\rm NL},\tilde{\theta}_{{\rm NL},i}
~\sim~{\rm Weibull}\!\big(\tilde{\alpha}_{\rm NL},\tilde{\theta}_{{\rm NL},i}\big),
\]
\[
\log \tilde{\theta}_{{\rm NL},i}
~=~ \tilde{\mathbf{Z}}_{{\rm NL},i}^{\top}\tilde{\boldsymbol{\beta}}_{\rm NL}
\;+\;\tilde{\phi}_3 \tilde{b}_{1i}\;+\;\tilde{\phi}_4 \tilde{b}_{2i}\;+\;\tilde{b}_{4i},
\]
yielding $\tilde{h}_{{\rm NL},i}(t)=\tilde{\alpha}_{\rm NL}\,\tilde{\theta}_{{\rm NL},i}\,t^{\tilde{\alpha}_{\rm NL}-1}$. The association $(\tilde{\phi}_3,\tilde{\phi}_4)$ maps the latent lesion process to NL risk, and $\tilde{b}_{4i}$ serves as the NL–specific frailty. Overall survival $\tilde{T}_{{\rm OS},i}$ depends on all four latent terms:
\[
\tilde{T}_{{\rm OS},i}\mid \tilde{\alpha}_{\rm OS},\tilde{\theta}_{{\rm OS},i}
~\sim~{\rm Weibull}\!\big(\tilde{\alpha}_{\rm OS},\tilde{\theta}_{{\rm OS},i}\big),
\]
\[
\log \tilde{\theta}_{{\rm OS},i}
~=~ \tilde{\mathbf{Z}}_{{\rm OS},i}^{\top}\tilde{\boldsymbol{\beta}}_{\rm OS}
\;+\;\tilde{\phi}_5 \tilde{b}_{1i}\;+\;\tilde{\phi}_6 \tilde{b}_{2i}\;+\;\tilde{\phi}_7 \tilde{b}_{3i}\;+\;\tilde{\phi}_8 \tilde{b}_{4i},
\]
so that $\tilde{h}_{{\rm OS},i}(t)=\tilde{\alpha}_{\rm OS}\,\tilde{\theta}_{{\rm OS},i}\,t^{\tilde{\alpha}_{\rm OS}-1}$. The coefficients $(\tilde{\phi}_5,\tilde{\phi}_6,\tilde{\phi}_7,\tilde{\phi}_8)$ quantify how the latent lesion level, its slope, and the two progression–component frailties translate to death risk. For OS prediction, we generate posterior predictive OS draws for each participant who is “still alive”.

\subsection{Prediction Models Based on Disease Progression as a Composite Endpoint}

\subsubsection{Copula Models for TTP and OS}
\label{sec:method_copula}
\begin{figure}
\centering
\subfloat[]{\includegraphics[width=4cm]{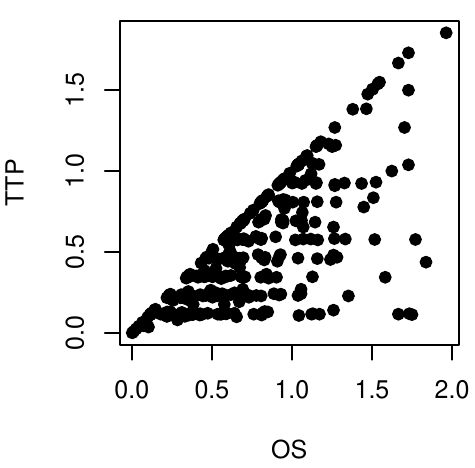}}\label{fig:scatter}
\subfloat[]{\includegraphics[width=4cm]{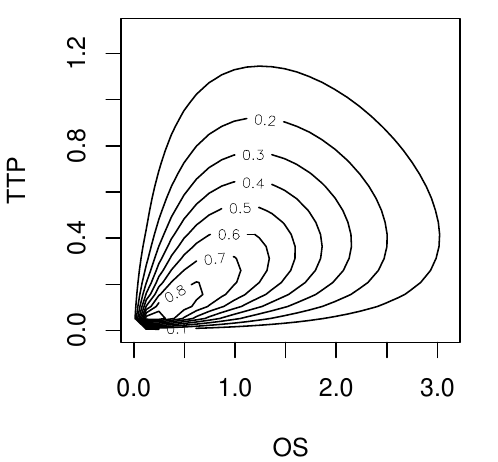}}\label{fig:contour}
\subfloat[]{\includegraphics[width=4cm]{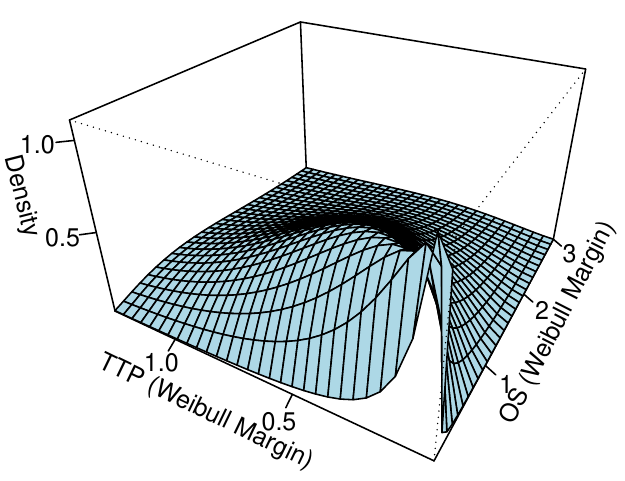}}\label{fig:surface}
\caption{(a) The scatterplot between time to progression (TTP) and overall survival (OS), (b) the contour plot of TTP vs. OS fitted with a Clayton copula, (c) the surface plot of TTP vs. OS fitted with a Clayton copula. \label{fig:copula}}
\end{figure}

Modeling TTP is similar to modeling PFS, but TTP excludes deaths from any cause. If a patient dies without documented disease progression, their TTP is censored at the last follow-up. Similar to the model defined in  \ref{sec:NT}, we use $\lambda_{\rm TTP}(t|\textbf{Z})$ to
denote the hazard function for TTP. Under the Cox
proportional hazards model, we have $\lambda_{\rm TTP}(t|\textbf{Z})=\lambda_{0, \rm TTP}(t)\mathrm{exp}(\textbf{Z}' \boldsymbol{\beta_{\rm TTP}})$, and the corresponding survival function is given by
$S_{\rm TTP}(t|\textbf{Z})=\mathrm{exp}\{-\gamma_{\rm TTP} t^{\alpha_{\rm TTP}}
\mathrm{exp}(\textbf{Z}' \boldsymbol{\beta_{\rm TTP}} )\}$, such that $\gamma_{\rm TTP}$ and $\alpha_{\rm TTP}$ are the scale and shape parameters
of the Weibull distribution, respectively. The Clayton model between TTP and OS is specified as
\begin{equation*}
S_1(t_{\rm TTP},
t_{\rm OS}|\textbf{Z})=\{S_{\rm TTP}(t_{\rm TTP}|\textbf{Z})^{-\eta_{\rm TTP}}
+S_{\rm OS}(t_{\rm OS}|\textbf{Z})^{-\eta_{\rm TTP}}-1\}^{-1/\eta_{\rm TTP}}, 
\end{equation*}
where $\eta_{\rm TTP}>0$ measures the correlation. Figure \ref{fig:copula} illustrates the Clayton copula fitted between TTP and OS.
The likelihood derivation can be found in supplementary materials. OS predictions are obtained using the PPD method described in Section \ref{sec:PPD}.

\subsubsection{Multi-State Model} \label{sec:multistate}
A multi-state model is another common approach for modeling the event history of participants in clinical trials \citep{andersen2002multi, meira2009multi, putter2007tutorial}, but currently, no literature has covered survival time prediction based on a multi-state model. Among the various multi-state models, the one that finds the widest application is the three-state illness-death model, which includes a singular immediate state denoting ``illness" (Supplementary Figure 3). In this paper, we employ the homogeneous semi-Markov assumption \citep{cox1977theory} for the multi-state model, which implies that the hazard of death after progression depends on the time since progression rather than time since randomization. Therefore, for every patient, we examine two distinct periods: the time between randomization and progression, and the duration from progression to death. Both of these intervals are treated as separate components and individually modeled. A third scenario occurs when a patient dies without experiencing progression.

For the multi-state model, transition probabilities represent the probabilities of transition from one state to another over a given time period. Here, we denote the transition probability as $P_{kl}(t_1, t_2)$ from time $t_1$ to $t_2$, where $k$, $l$ describes the state with $k \in {0, 1, 2}$ and $l \in {0, 1, 2}$. The expressions of the transition probabilities are given by \citep{meira2009multi}:
\begin{eqnarray*}
P_{00}(t_1, t_2) &=& S_0(t_2 - t_1) = exp{(-\prod_{01}(t_2 - t_1) - \prod_{02}(t_2 - t_1))},\\
P_{11}(t_1, t_2) &=& S_1(t_2 - t_1) = exp{(-\prod_{12}(t_2 - t_1))},\\
P_{12}(t_1, t_2) &=& S_1(t_2 - t_1) = \int_{t_1}^{t_2}P_{11}(t_1, u) \pi_{12}(u;\mathcal{F}_u)P_{22}(u, t_2)du,
\end{eqnarray*}
where $\prod_{kl}(t_1, t_2)=\int_{t_1}^{t_2}\pi_{kl}(t, \mathcal{F}_t)dt$ is the cumulative transition intensity between states $k$ and $l$, where $k \leq l$. If we consider the transition intensities $\pi_{kl}(t; \mathcal{F}_t)$ to follow Weibull distributions, they can be expressed as
$\pi_{01}(t)=\alpha\left(1/\gamma_{01}\right)^\alpha t^{\alpha - 1},$
$\pi_{02}(t)=\alpha\left(1/\gamma_{02}\right)^\alpha t^{\alpha - 1},$
and $\pi_{12}(s)=\alpha\left(1/\gamma_{12}\right)^\alpha s^{\alpha - 1},$
where $\gamma_{01}, \gamma_{02}$ and $\gamma_{03}$ are the scale parameters, $\alpha$ is the shape parameter, $t$ refers to time since randomization and $s$ refers to time since progression. $\gamma_{01}, \gamma_{02}$ and $\gamma_{03}$ for each patient are further defined as $\gamma_{01,i} = \Tilde{\gamma}_{01}exp(\boldsymbol{\beta}_{01}\boldsymbol{X}_i)$, $\gamma_{02,i} = \Tilde{\gamma}_{02}exp(\boldsymbol{\beta}_{02}\boldsymbol{X}_i)$, and $\gamma_{12,i} = \Tilde{\gamma}_{12}exp(\boldsymbol{\beta}_{12}\boldsymbol{X}_i)$, respectively, where $i \in N$, $\boldsymbol{X}_i$ is a vector of covariates, and $\boldsymbol{\beta}_{01}$, $\boldsymbol{\beta}_{02}$ and $\boldsymbol{\beta}_{12}$ are regression coefficients. $\Tilde{\gamma}_{01}$, $\Tilde{\gamma}_{02}$, and $\Tilde{\gamma}_{12}$ are the baseline hazards. To guarantee the convergence of MCMC, a uniform shape parameter $\alpha$ is employed for all three Weibull functions. The multi-state model's likelihood construction is provided in supplementary materials. OS predictions are obtained using the PPD method described in Section \ref{sec:PPD}.

\subsection{Marginal Weibull Baseline Hazard Model of OS}\label{sec:method_standard}

Most commonly, PFS is modeled using a Kaplan-Meier curve or Cox regression. For example, a simple Weibull baseline hazard model has the form
$\lambda(t)=\alpha \gamma t^{\alpha-1} \exp(\textbf{Z}' \boldsymbol{\beta})$,
where $\textbf{Z}_i$ is the covariate matrix for modeling PFS, $\alpha$ and $\gamma$ are the shape and scale parameters of the Weibull distribution, and $\boldsymbol{\beta}$ is a vector of coefficients. Note that a standard analytic method does not consider the dynamics of the component processes of PFS and does not take random effects into account.

\subsection{Prior Specification}

In this section, we provide recommendations for specifying priors for the parameters in the models discussed in Section \ref{sec:method}. Our aim is to promote the use of generalizable, weakly informative, or non-informative priors for all model parameters to ensure robust and interpretable results. We emphasize the importance of selecting appropriate priors to maintain parameter identifiability.

\subsubsection{General Guidelines}

For the coefficient vectors $\boldsymbol{\beta}$ and $\tilde{\boldsymbol{\phi}}$, we recommend weakly informative normal priors centered at 0. We suggest weakly informative exponential distributions for the shape parameters ($\alpha$) in the Weibull distribution. All random effect parameters ($b$) should follow weakly informative normal distributions, denoted as $N(0, \sigma_{b_k}^2)$, where $k$ stands for different random variables under different models. Weakly informative half-normal priors are recommended for the standard deviations ($\sigma_{b_k}$) associated with the random effects. For all Clayton copula parameters ($\eta$), we recommend using weakly informative exponential distributions. These priors can accommodate a wide range of plausible values.

In the model assessing the association between target lesion and OS, which is a component of baveJM, we suggest the following priors. For the parameter $\lambda$, a weakly informative normal distribution with a mean of 0 is recommended. The correlation coefficient ($\rho$) should have a non-informative prior, such as $Unif(-1, 1)$, to avoid biasing the estimation. Additionally, we assume that all models are equally weighted before fitting BMA, that is, $P(M_q) = 1/Q$, for all $q = 1,...,Q$.

In the multi-state model, we advise using weakly informative half-normal priors for all $\Tilde{\gamma}$ parameters. These priors allow flexibility while constraining extreme values.

\section{Simulations}
\label{sec:simulation}

\subsection{Design}
We performed extensive simulations to evaluate the performance of the proposed baveJM approach. All simulated datasets mimic the structure of the RCC trial data. Specifically, we assume a trial with 400 subjects and 25 visits. Longitudinal measurements of target lesion and new lesion status are recorded at each visit. For simplicity purposes, non-target lesion status is not included. Survival status is updated whenever deaths occur. Follow-up visits occur every two months within the first two years, then change to every three months until the end of the fifth year. We censor the subject's subsequent target lesion measurements and new lesion status whenever that subject dies. 

We simulate under four scenarios: 1) OS is independent of both target lesion measurements and time to appearance of new lesion development; 2) OS is correlated only with time to appearance of new lesion; 3) OS is correlated only with target lesion measurements; and 4) OS is correlated with both target lesion measurements and time to appearance of new lesion development. In all scenarios, both target lesion measurements and new lesion statuses are simulated randomly and independently. However, OS is generated under varying conditions. Specifically, we simulate target lesion measurements using a linear mixed-effects model that incorporates time and treatment as fixed effects, alongside a patient-level random intercept and a random time effect. Furthermore, the time to appearance of new lesion is modeled using a Weibull baseline hazard function. In scenario 1, OS is randomly and independently simulated based on a Weibull baseline hazard function. In scenario 2, OS is conditionally simulated using a bivariate Clayton copula model, linking OS with time to appearance of new lesion. In scenario 3, OS is conditionally simulated based on target lesion measurements. Here, OS is still simulated using a Weibull baseline hazard function, but with an assumed negative correlation between quantiles of OS and larger tumor measurements. In scenario 4, OS is conditionally simulated using both mechanisms: it is linked to time to appearance of a new lesion through a stronger bivariate Clayton copula, and it also depends on the subject’s average target-lesion size (larger averages imply shorter OS). For each scenario, 1000 datasets were generated. To streamline the simulation process, only treatment was included in the covariate matrices.

We compare predictions under five models: our proposed baveJM approach, SPJM, a copula model for TTP and OS, a multi-state model, and a marginal Weibull baseline hazard model of OS. 

\begin{figure}
\centering
\subfloat[]{\includegraphics[width=7cm]{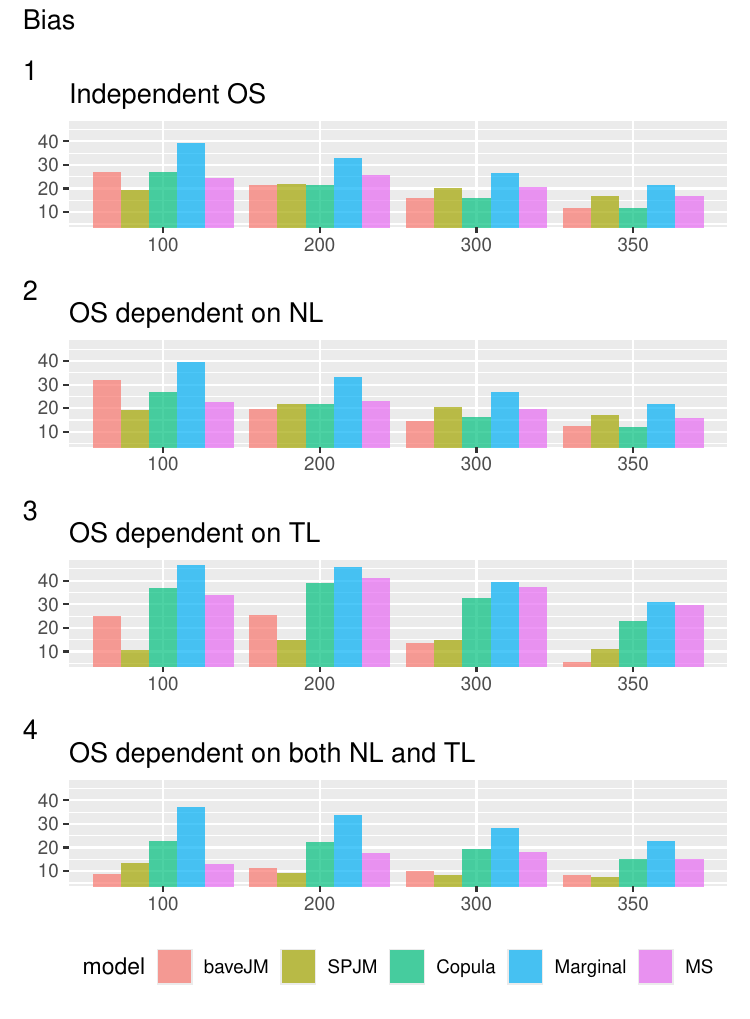}\label{fig:bias}}
\subfloat[]{\includegraphics[width=7cm]{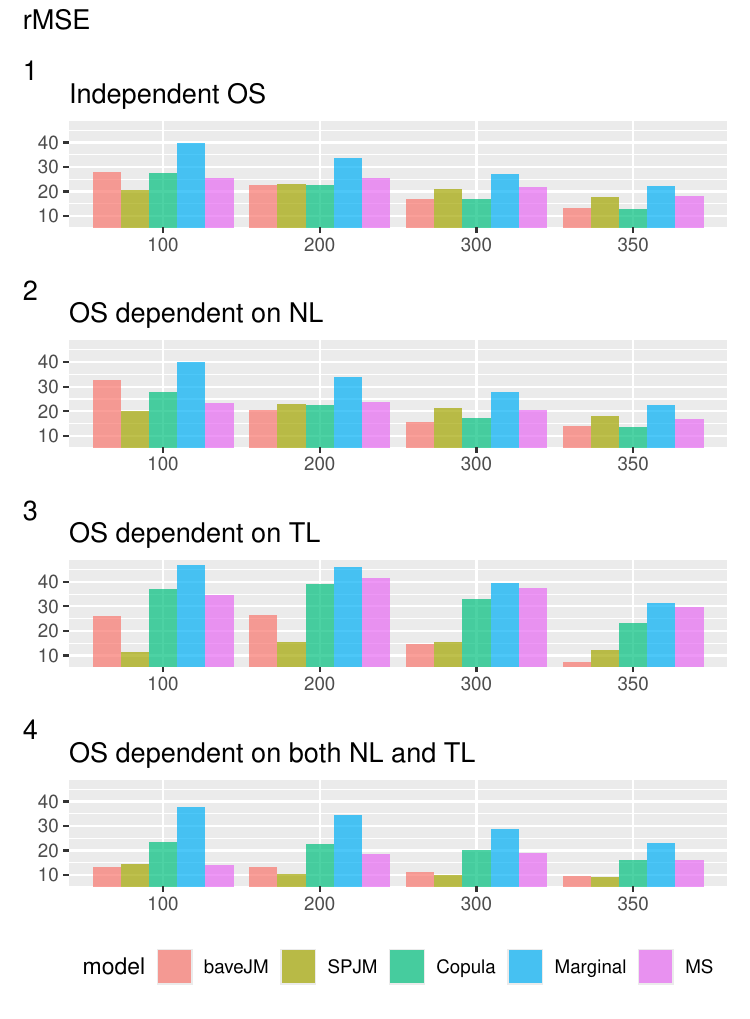}\label{fig:rmse}}
\caption{The bias (a) and root mean squared error (rMSE) (b) of the last (400th) death date predictors of five models: the multivariate joint modeling approach (baveJM), a shared-parameter random-effect joint model (SPJM), a copula model between TTP and OS (Copula), the marginal Weibull baseline hazard model of OS (Marginal), and a multi-state model (MS), under four OS scenarios. The $x$-axis denotes the number of deaths at the cutoff. The $y$-axis denotes the number of months. \label{fig:result}}
\end{figure}

To emulate real trial scenarios and assess the prediction performance of our proposed method, we generated snapshots of the data. For each simulated dataset, we took snapshots of what the data would have looked like if only 100, 200, and 300 death events had occurred. Then, we predicted the time of the last (400th) death event under each snapshot dataset. Note that predictions may be performed at any time during the trial, and 100, 200, and 300 were selected for illustration purposes. Predictions from the five methods were compared. Finally, we calculated each predictor's coverage rate (CR), root-mean-squared error (rMSE), bias, and width of the 90\% credible interval (CI). The 90\% CI is the narrowest interval that includes 90\% of the posterior distribution of the predictor. All simulations are done in R. The code is provided in the supplementary materials.

\subsection{Results}

The bias and rMSE of the predicted time of the last death are shown in Figures \ref{fig:bias} and \ref{fig:rmse}, and the corresponding CR and CI width are shown in Supplementary Table 1. The weights of all submodels of baveJM under each simulation scenario are shown in Supplementary Figure 4.

Under scenario 1, where OS is generated independently, the bias and rMSE of all predictors decrease as more deaths are observed. Overall, the marginal model performs worse than the other four methods. This is because the marginal Weibull uses only OS and, under substantial censoring, tends to anchor predictions near the upper end of the observed follow-up instead of borrowing information from progression components. Given that baveJM placed essentially all its weight on the NL–OS copula model, and the Clayton copula collapses to independence when its dependence parameter is near zero, it is natural that the copula model and baveJM exhibit similar bias and rMSE. SPJM outperforms baveJM and the copula model at 100 deaths but its improvement stalls and its CR declines as more deaths accrue, consistent with over-conditioning on shared random effects under a misspecified target lesion and OS link that yields overconfident, too-narrow predictive intervals. Considering all metrics, baveJM and the copula model are the best-performing methods in scenario 1.

In Scenario 2, OS and the time to the appearance of a new lesion are generated jointly using a bivariate Clayton copula, so OS depends on NL through the dependence structure. As the number of observed deaths increases, bias and rMSE decrease for all methods. The marginal OS model continues to lag because it ignores progression information, foregoing efficiency and calibration gains available from NL–OS dependence. baveJM and the copula model show very similar bias and rMSE, as baveJM places most of its weight on the NL–OS copula component under this scenario. The multi-state model improves relative to Scenario 1 because it can leverage the presence of NL–OS dependence. SPJM’s performance largely mirrors Scenario 1: it benefits early from borrowing the longitudinal target lesion process. Although SPJM includes an NL–OS shared-effect term, its gaussian-type dependence on the log scale cannot capture the Clayton lower-tail dependence used to generate Scenario 2. Consequently, for Scenario 2 we recommend baveJM or the copula model.

In Scenario 3, where OS depends on the target lesion trajectory, baveJM and SPJM, the only methods that explicitly include a target lesion and OS link, clearly outperform the other three. With 100 and 200 deaths, SPJM has slightly lower bias and rMSE, likely because dense target lesion measurements yield precise subject-level effects that SPJM maps directly onto the OS scale, reducing variance early. As more deaths accrue (300 and 350), baveJM reallocates most of its weight to the target lesion and OS joint component and overtakes SPJM. Unlike Scenarios 1 and 2, the multi-state model does not improve monotonically, since it leverages progression states rather than the target lesion trajectory that actually drives OS here. SPJM has the widest intervals. It carries over uncertainty from the target lesion measurements and slope terms and also estimates extra terms (e.g., NT/NL frailties) that do not matter here, which adds noise. By contrast, as data accumulate, baveJM puts most of its weight on the single target lesion and OS submodel, so its intervals are tighter in comparison.

In Scenario 4, OS depends on both the target lesion measurements and the time to appearance of a new lesion. Bias and rMSE tend to improve with more deaths, but not monotonically across methods. The marginal OS model still lags because it ignores both target lesion measurements and new lesion information. baveJM and SPJM outperform the copula model, since they include a target lesion and OS link in addition to modeling NL–OS dependence. baveJM and SPJM perform similarly overall, with SPJM marginally ahead at some death counts, likely reflecting its direct target lesion and OS link (i.e., no averaging penalty) when target lesion measurements are highly informative. 

According to Figure \ref{fig:result}, across all scenarios, the proposed baveJM approach performs consistently well and is the most reliable overall. When OS is independent of target lesion measurements (Scenario 1) or driven by NL–OS dependence (Scenario 2), baveJM closely tracks the copula model on bias and rMSE by placing most of its weight on the NL–OS component; when OS is driven by target lesion (Scenario 3), baveJM shifts weight to the target lesion and OS submodel and catches up to—and with more events, surpasses—SPJM; when both signals are present (Scenario 4), baveJM remains on par with SPJM and outperforms the copula model by incorporating the target lesion effect. SPJM tends to show strong early point accuracy when target lesion is informative, but can be less well-calibrated when the target lesion and OS link is weak or absent. baveJM’s adaptive weighting yields competitive rMSE while maintaining robustness across the different data-generating scenarios.

\begin{figure}
\centering
\subfloat[]
{\includegraphics[width=12cm]{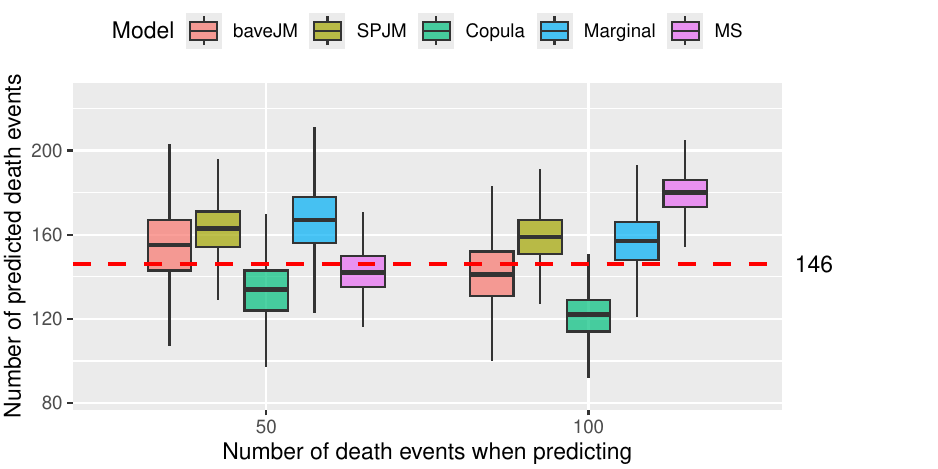}\label{fig:primaryanalysis}}
\hfill
\subfloat[]
    {\includegraphics[width=12cm]{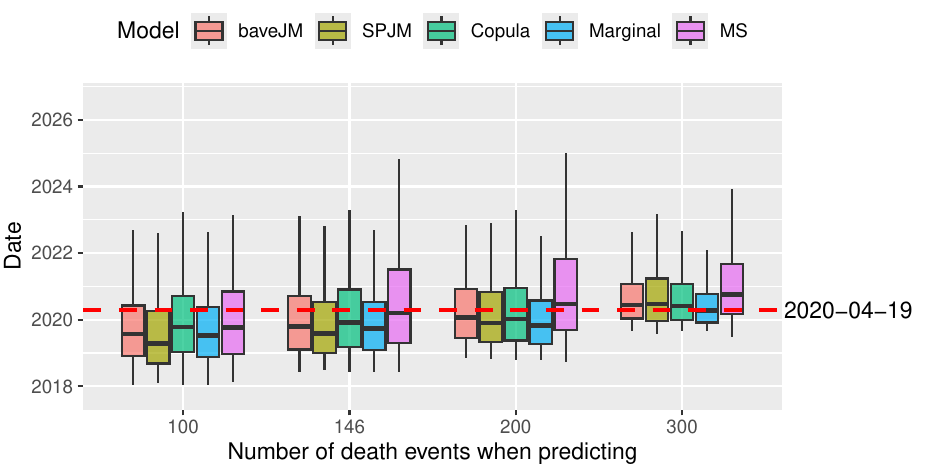}\label{fig:casestudy}}
\hfill
\subfloat[]
{\includegraphics[width=12cm]{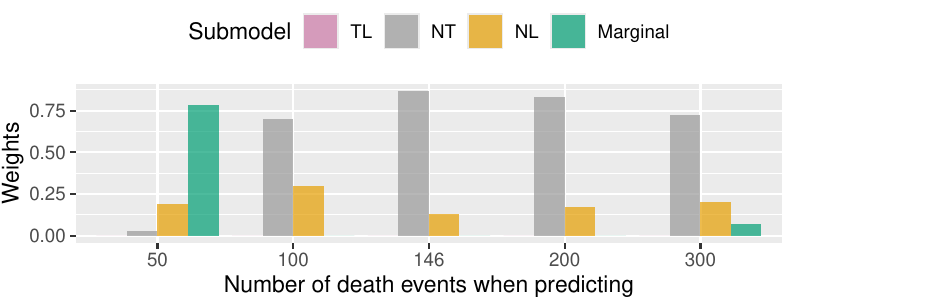}\label{fig:weights}}
\caption{\footnotesize (a) Boxplots of the predicted number of deaths at primary analysis when 50 or 100 deaths are observed in the trial, with the true number of deaths being ``146".(b) Boxplots of the posterior distributions of the last (341st) death date by baveJM, SPJM, the copula model between TTP and OS, the marginal Weibull baseline hazard model of OS, and the multi-state model. The date ``2020-04-19" is the true last (341st) death date. (c) The final weights of each submodel under baveJM.}
\end{figure}

\section{Analysis of the Renal Cell Carcinoma Trial Data}
\label{sec:caseanalysis}
We return to the RCC trial dataset introduced in Section \ref{sec:example}. To evaluate the model's performance, we considered the trial data after 341 deaths were observed. The objective is to compare the predicted timing of the 341st death using different models described in Section \ref{sec:method} with the actual observed timing. The model's performance was further evaluated by comparing the number of deaths predicted by different models with the observed number of deaths at the PFS final analyses. For this exercise, we generated snapshots of the trial data at the 100th, 146th (the time of the primary analysis), 200th, and 300th death and compared predictions under five models: baveJM, SPJM, a copula model for TTP and OS, a multi-state model, and a marginal model of OS. In addition to timing, we are interested in calculating the likelihood of demonstrating statistically significant OS in favor of the experimental drug at the updated OS analysis after 341 deaths.

We included \textit{gender, age, nephrectomy at baseline, Heng prognostic criteria at baseline, and ECOG Performance Status} into the covariate matrix $\textbf{X}_i$ and $\textbf{Z}_i$. Then, we fit each model using two MCMC chains, each with at least 10,000 MCMC iterations, in addition to 1,000 burn-in and 8,000 adaptation iterations. Convergence diagnostic tests, trace plots, and autocorrelation were investigated to ensure convergence. Figure \ref{fig:primaryanalysis} displays the predicted number of deaths at the time of the primary analysis. Figure \ref{fig:casestudy} presents boxplots of the predictive distributions for the predictors of the last (341st) death date from all tested models compared to the true last (341st) death date. The median values of the boxplots serve as point predictions. Figure \ref{fig:weights} displays the posterior weights of all submodels for each dataset based on the specified number of deaths. In this case study, baveJM, SPJM, and the copula model yield closely clustered central forecasts for the primary analysis death count, whereas the marginal and MS models exhibit greater dispersion. For the predicted date of the last (341st) death, baveJM and the copula model remain well-centered with comparatively tight dispersion across looks (100, 146, 200, and 300 deaths), and SPJM is broadly comparable. The marginal and MS models are substantially more variable. The submodel-weight plot indicates that baveJM does not commit to a single mechanism. Instead, it adaptively reallocates weight among TL, NT, NL, and marginal components as additional deaths accrue. The posterior odds favor the submodel capturing the association between non-target lesions and OS, consistent with current disease knowledge. Taken together, these figures indicate that baveJM is the most robust across analyses, the copula model generally matches baveJM when NL/TTP carries signal, SPJM remains competitive without a persistent advantage, and the MS and marginal models display greater uncertainty in this case study.

Table \ref{tab:casestudytable} presents the bias and rMSE of the predicted OS from different models when 100, 146, 200, or 300 death events are observed. In this case study, baveJM performs similarly to SPJM and the copula model. Additionally, baveJM exhibits smaller bias and rMSEs compared to the marginal model under all scenarios. Overall, baveJM is consistently competitive, often best or near best on rMSE, while the copula model catches up or slightly leads at 300 death events. These patterns are consistent with Figures~\ref{fig:primaryanalysis}–\ref{fig:weights}, which show stable point predictions for the last death date and adaptive reweighting of submodels as information accrues. 

\begin{figure}
    \centering
    \subfloat[]{\includegraphics[width=7cm]{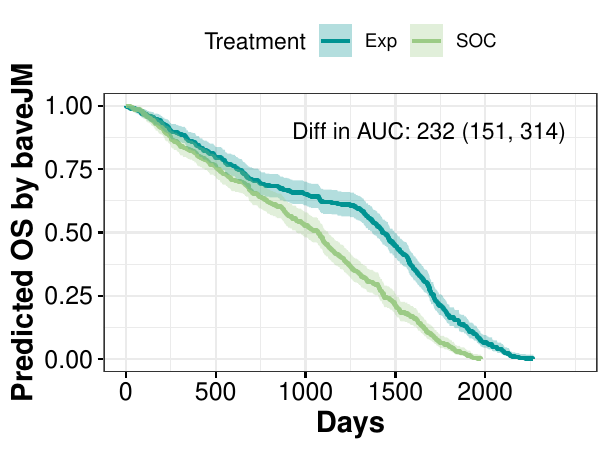}}\label{fig:JMprediction}
    \subfloat[]{\includegraphics[width=7cm]{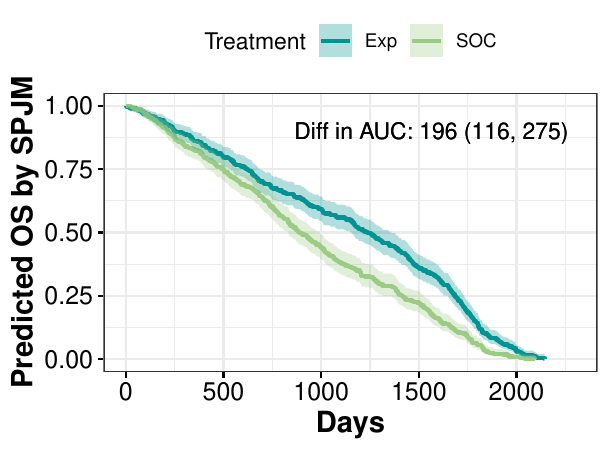}}\label{fig:FullJMprediction}
    \subfloat[]
    {\includegraphics[width=7cm]{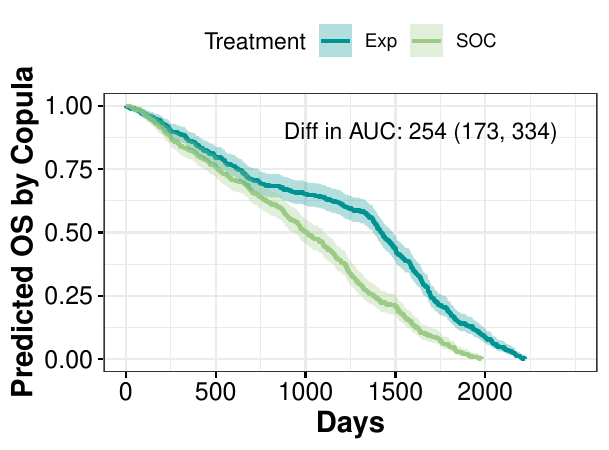}}\label{fig:Copulaprediction}
    \subfloat[]{\includegraphics[width=7cm]{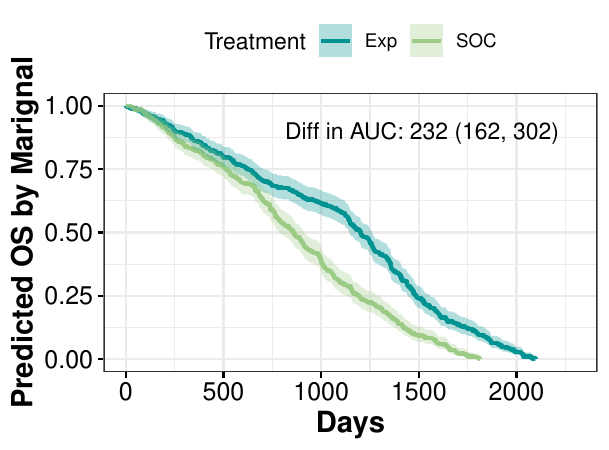}}\label{fig:Marginalprediction}
    \subfloat[]{\includegraphics[width=7cm]{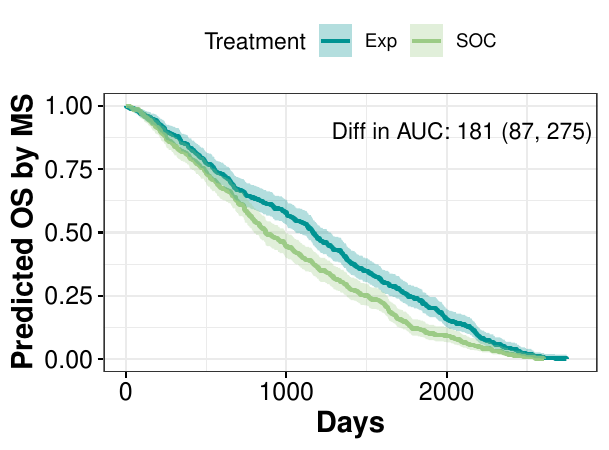}}\label{fig:MSprediction}
    \caption{Kaplan-Meier plots of the observed and predicted overall survival by baveJM (a), SPJM (b), the copula model (c), the marginal model (d), and the multi-state model (e). The difference in the area under the curve (AUC) and its credible intervals for the experimental drug (Exp) and the standard of care (SOC) are also included.}
    \label{fig:predictionKM}
\end{figure}

Finally, We used five models to forecast the death time of patients who were still alive after 341 deaths were observed. Subsequently, we generated predicted Kaplan-Meier plots for each model (Figure \ref{fig:predictionKM}) and estimated the potential gain or loss in life expectancy by calculating the difference in the area under the curve between the experimental drug and the standard of care \citep{pak2017interpretability, survRM2}. The results from Figure \ref{fig:predictionKM} indicate that all models consistently project a substantial improvement in life years. In addition, baveJM, SPJM and the copula model, whose reliability was previously established in this case study, all forecast a at least 6.5-month gain in life years for patients who receive the experimental drug, as opposed to the standard of care.

\begin{table}
\caption{Bias and rMSE of the predictors of baveJM, SPJM, copula model between TTP and OS, marginal model, and multi-state model (MS). \label{tab:casestudytable}}
\begin{center}
\begin{tabular}{ccccccccc}
\hline
\multicolumn{2}{c}{No. of events} & \multirow{2}{*}{Metric} & \multirow{2}{*}{baveJM} & \multirow{2}{*}{SPJM} & \multirow{2}{*}{Copula} & \multirow{2}{*}{Marginal} & \multirow{2}{*}{MS}\\ 
Observed & Predicted & & & & & \\\hline
\multirow{2}{*}{100} & \multirow{2}{*}{41} & Bias & 11.35 & 9.86 & 9.23 & 12.75  & 3.46 \\
&& rMSE & 12.08 & 10.74 & 10.25 & 13.44  & 8.24 \\ \hline
\multirow{2}{*}{146} & \multirow{2}{*}{41} & Bias & 6.88 & 7.18 & 7.41 & 10.90  & 7.23 \\
&& rMSE & 8.36 & 8.92 & 8.97 & 12.03 & 10.14 \\ \hline
\multirow{2}{*}{200} & \multirow{2}{*}{41} & Bias & 4.23 & 5.02 & 4.77 & 8.57 & 6.56 \\
&& rMSE & 7.24 & 8.14 & 7.52 & 10.32 & 9.87 \\  \hline
\multirow{2}{*}{300} & \multirow{2}{*}{41} & Bias & 2.91 & 2.59 & 2.99 & 5.11 & 5.54  \\
&& rMSE & 8.57 & 8.80 & 8.46 & 9.26 & 11.24  \\ \hline
\end{tabular}
\end{center}
\end{table}

\section{Discussion}
\label{sec:discussion}

We were motivated by an advanced RCC clinical trial to demonstrate how real-time OS predictions from joint models with different components of PFS can be dynamically combined via BMA. The proposed baveJM delivers reliable estimates of the time of the $n$th death and the probability of demonstrating an OS benefit after observing $n$ deaths, leveraging all available RECIST 1.1 tumor assessments. These capabilities are useful for trial planning and for informing end-of-life care. In both the case study and the simulation scenarios, baveJM provides robust, well-calibrated predictions, adapting to whichever progression component carries the strongest signal. Although developed in RCC, the framework is directly applicable to other solid tumors.

baveJM is flexible and can accommodate linear longitudinal models with time-dependent covariates or nonlinear mixed-effects structures. BMA optimally pools evidence across submodels but can be computationally intensive. Because the covariates entering each submodel need not be identical, it is advantageous to pre-select covariates for each component using domain expertise or principled variable-selection procedures, which can reduce dimensionality and training time.

The copula model linking time to progression (e.g., new lesion or TTP) and OS emerges as a strong performer. When TTP and OS are well described by Weibull marginals with random effects, the copula model typically outperforms a marginal OS-only model. As a practical option to reduce computation, fitting a copula model with random effects alone can deliver accurate predictions when progression–survival dependence is present. Conversely, omitting random effects can yield overly heavy-tailed predictions in heterogeneous populations.

The SPJM is competitive across analyses, particularly when the target lesion process is informative. With frequent  target lesion assessments, subject-level target lesion effects are estimated precisely and a direct target lesion and OS link in SPJM can reduce variance and yield strong early point accuracy. However, when the target lesion and OS association is weak or absent, SPJM can be less well calibrated because it continues to load shared effects onto the OS scale. 

For a marginal OS model to perform better, it would be beneficial to incorporate more predictive covariates or use a model offering greater flexibility than the Weibull baseline hazard model, such as piecewise exponential models. Similarly, to improve the performance of the multi-state model, we could replace the Weibull baseline hazard model with more flexible models.

During the simulation studies, the data for non-target lesions were not simulated, yet they were included in the example data analysis. This does not undermine the outcome of our simulation studies, given that both the non-target lesion and new lesion were modeled using the exact same joint model, and each component of PFS was modeled separately. If the non-target lesion data had been included in the simulation dataset, we would have anticipated longer training times for baveJM without observing any relative improvement in performance.

Apart from applying the same model weights to all patients in the dataset, BMA can potentially provide personalized model weights, where each patient may have different weights for each model. This could further improve the prediction accuracy. Future work can explore how to implement personalized model weights in a time-efficient manner in this study setting.

\bibliographystyle{imsart-nameyear} 
\bibliography{reference}       

\end{document}


\bibliographystyle{agsm}

\def\spacingset#1{\renewcommand{\baselinestretch}%
{#1}\small\normalsize} \spacingset{1}


\if1\blind
{
  \title{\bf Supplementary Materials for \textit{Dynamic Prediction of Milestones for Survival Endpoints in Metastatic Solid Tumor Cancer Clinical Trials}}
  \author{Sidi Wang\\
    Department of Biostatistics, University of Michigan\\
    Kelley M Kidwell \\
    Department of Biostatistics, University of Michigan\\
    Bo Huang \\
    Pfizer Inc., New York, NY, U.S.A. \\
    and \\
    Satrajit Roychoudhury \\
    Pfizer Inc., New York, NY, U.S.A.}
  \maketitle
} \fi

\if0\blind
{
  \bigskip
  \bigskip
  \bigskip
  \begin{center}
    {\LARGE\bf Improving prediction of overall survival using joint modeling for oncology studies with solid tumor response}
\end{center}
  \medskip
} \fi

\bigskip

\noindent%

\vfill

\spacingset{1.9} 
\newpage
\begin{figure}[H]
    \centering
    \subfloat[]{\includegraphics[width=0.45\textwidth]{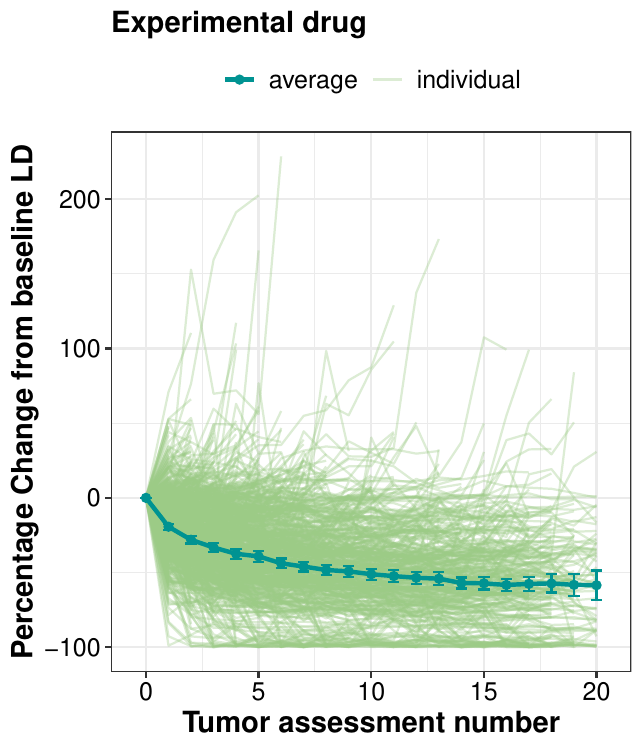}\label{fig:profile_LD_exp}}
    \hfill
    \subfloat[]{\includegraphics[width=0.45\textwidth]{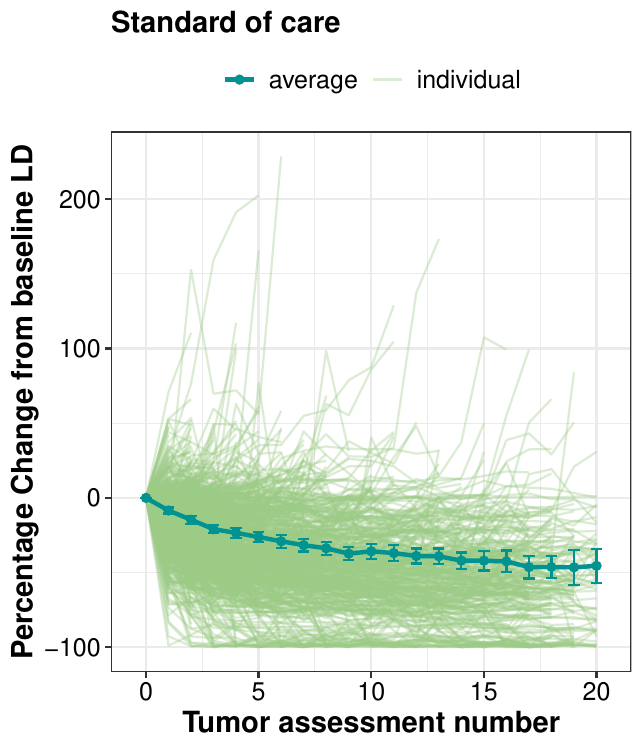}\label{fig:profile_LD_soc}}
    \caption{Profile plot of the sum of the longest diameter of target lesions}
    \label{fig:profile_LD}
\end{figure}

\newpage
\begin{figure}[H]
    \centering
    \subfloat[]{\includegraphics[width=0.45\textwidth]{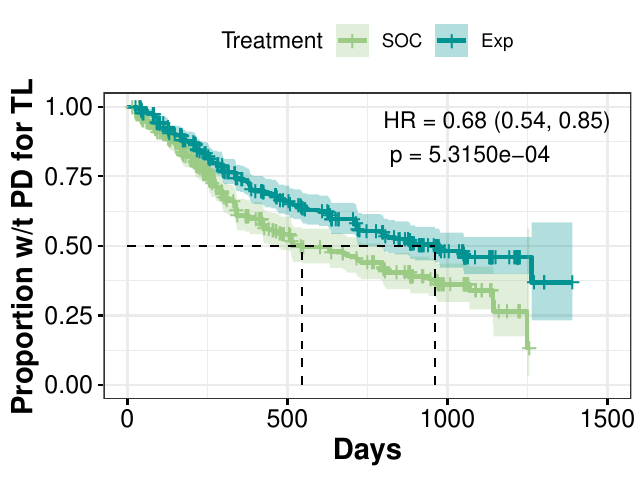}\label{fig:TL_KMplot}}
    \subfloat[]{\includegraphics[width=0.45\textwidth]{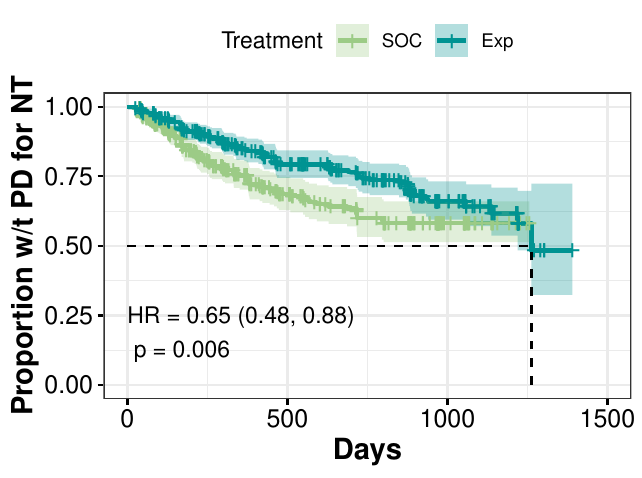}\label{fig:NT_KMplot}}
    \hfill
    \subfloat[]{\includegraphics[width=0.45\textwidth]{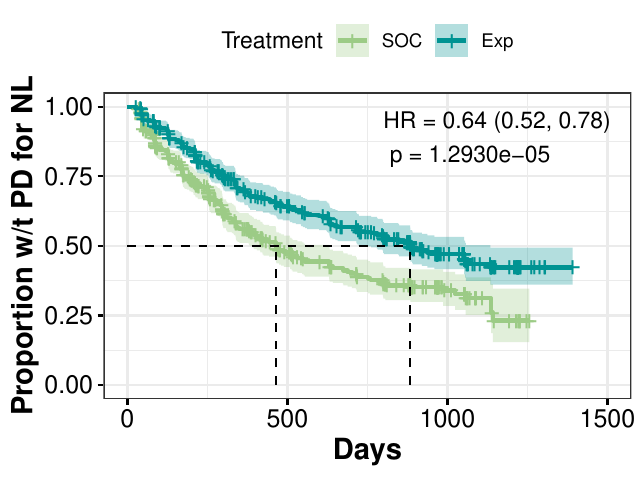}\label{fig:NL_KMplot}}
    \subfloat[]{\includegraphics[width=0.45\textwidth]{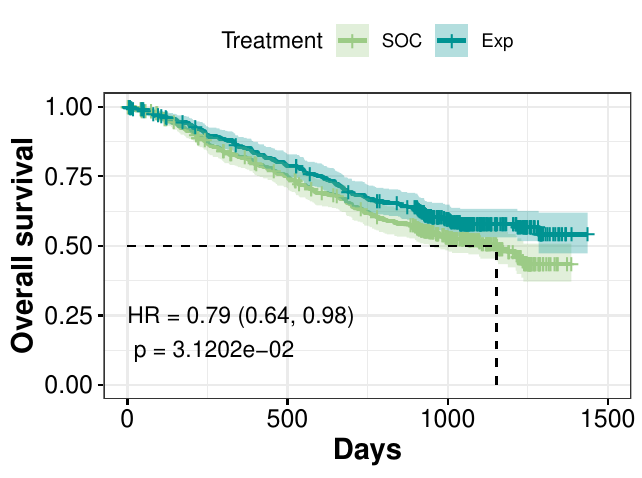}\label{fig:OS_KMplot}}
    \hfill
    \subfloat[]{\includegraphics[width=0.45\textwidth]{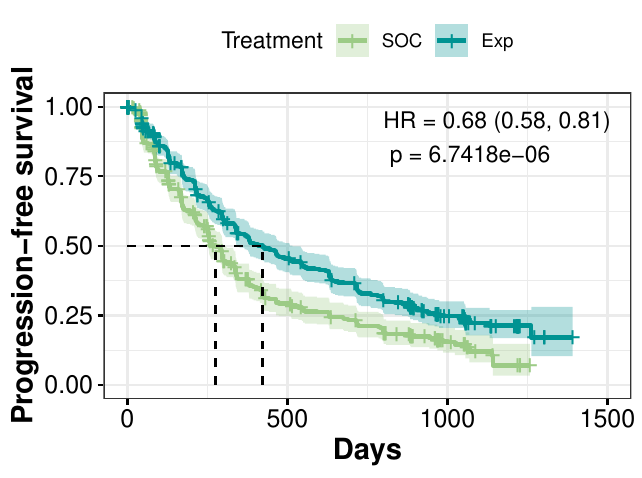}\label{fig:PFS_KMplot}}
    \caption{Kaplan-Meier plots at the time of updated OS analysis considering target lesions progressive disease (PD) (a), non-target lesions PD (b), new lesions PD (c), overall survival (d), and progression-free survival (e).}
\end{figure}


\newpage
\section{Likelihood derivation for the joint model between target lesion and OS}
The marginal distribution of the observed measurements $\mu$ is easily obtained. The likelihood for the observed data can be factorized as the product of this marginal distribution and the conditional distribution of OS, given the observed values of $\mu$. Let $\boldsymbol{\theta}_1$ denote the combined vector of unknown parameters. Conditional on lesion measurements $\mu$, OS is independent of these measurements $\mu$, so we can write the likelihood $L=L(\boldsymbol{\theta}_1, \mu, OS)$ as
$L=L_{\mu}(\boldsymbol{\theta}_1,\mu)  \times L_{OS| \mu}(\boldsymbol{\theta}_1,OS|\mu)
$,
where $L_{\mu}(\boldsymbol{\theta}_1,\mu)$ is of standard form corresponding to the marginal multivariate normal distribution of $\mu$ and
$$
L_{OS|\mu}(\boldsymbol{\theta}_1, OS|\mu)=\prod_i \bigg\{ \bigg[\lambda_0 (t) \exp(\textbf{Z}_i' \boldsymbol{\beta}_{\rm OS}+\lambda \mu_{i}(t)) \bigg]^{\delta_{OS,i}} 
$$
$$\times \exp \bigg[ -\int_0^{y_{OS,i}} \lambda_0(t) \exp(\textbf{Z}_i' \boldsymbol{\beta}_{\rm OS}+\lambda \mu_{i}(t))dt\bigg]\bigg\},
$$
Here, $y_{OS,i}=\min(t_{OS,i},c_{OS,i})$ and $\delta_{OS,i}=I(t_{OS,i}\leq c_{OS,i})$. $c_{OS,i}$ is the censoring time for the $i$th participant and $t_{OS,i}$ is the true event time.

\newpage
\section{Likelihood derivation for the copula model of the non-target lesion and OS}
For non-target lesion, we define $y_{\rm NT}=\mathrm{min}(t_{\rm NT}, c_{\rm NT})$ and
$\delta_{\rm NT}=I(t_{\rm NT} \le c_{\rm NT})$,
where $c_{\rm NT}$ and $I(\cdot)$ denote the censoring time and the indicator function,
respectively; and define $y_{\rm OS}$ and $\delta_{\rm OS}$ similarly for OS.

Depending on the censoring pattern, the observed data for the $i$th participant falls into one of the following mutually exclusive cases: 1) both $t_{NT}$ and $t_{OS}$ are observed ($\delta_{NT}$=1, $\delta_{OS}$=1), 2) $t_{NT}$ is observed and $t_{OS}$ is censored ($\delta_{NT}$=1, $\delta_{OS}$=0), 3) $t_{NT}$ is censored and $t_{OS}$ is observed ($\delta_{NT}$=0, $\delta_{OS}$=1), and 4) both $t_{NT}$ and $t_{OS}$ are censored ($\delta_{NT}$=0, $\delta_{OS}$=0). Based on these four scenarios, we derive the likelihood of the copula model. 

Let
$\boldsymbol{\theta}_2=(\boldsymbol{\beta_{\rm NT}}, \alpha_{\rm NT}, \lambda_{\rm NT}, \boldsymbol{\beta_{\rm OSNT}},
\alpha_{\rm OSNT}, \lambda_{\rm OS}, \eta_{\rm NT})$,  then the
likelihood for the $i$th patient with ${\rm Data}_i=(y_{\rm NT}, y_{\rm OS},
\delta_{\rm NT}, \delta_{\rm OS}, \textbf{Z})_i$ is given by
$$
L(\boldsymbol{\theta}_2|{\rm Data}_i)
=L_1^{\delta_{\rm NT}\delta_{\rm OS}}
L_2^{\delta_{\rm NT}(1-\delta_{\rm OS})}L_3^{(1-\delta_{\rm NT})\delta_{\rm OS}}
L_4^{(1-\delta_{\rm NT})(1-\delta_{\rm OS})}, 
$$

where
$$
L_1=\frac{\partial^2\, S_1(t_{\rm NT}, t_{\rm OS}|\textbf{Z})}{\partial t_{\rm NT} \partial t_{\rm OS}}\big|_{t_{\rm NT}=y_{\rm NT},t_{\rm OS}=y_{\rm OS}}
$$
$$
=\bigg(\eta_{\rm NT}+1\bigg)\bigg(\mathrm{exp}\{-\gamma_{\rm NT} y_{\rm NT}^{\alpha_{\rm NT}}
\mathrm{exp}(\textbf{Z}' \boldsymbol{\beta_{\rm NT}} )\} \mathrm{exp}\{-\gamma_{\rm OS} y_{\rm OS}^{\alpha_{\rm OSNT}}
\mathrm{exp}(\textbf{Z}' \boldsymbol{\beta_{\rm OSNT}} )\}\bigg)^{-(\eta_{\rm NT}+1)}
$$
$$
\times \bigg(\mathrm{exp}\{-\gamma_{\rm NT} y_{\rm NT}^{\alpha_{\rm NT}}
\mathrm{exp}(\textbf{Z}' \boldsymbol{\beta_{\rm NT}} )\}^{-\eta_{\rm NT}}+\mathrm{exp}\{-\gamma_{\rm OS} y_{\rm OS}^{\alpha_{\rm OSNT}}
\mathrm{exp}(\textbf{Z}' \boldsymbol{\beta_{\rm OSNT}} )\}^{-\eta_{\rm NT}}-1\bigg)^{-\frac{2\eta_{\rm NT}+1}{\eta_{\rm NT}}}
$$
$$
\times \gamma_{\rm NT} \alpha_{\rm NT} y_{\rm NT}^{{\alpha_{\rm NT} -1}} \exp(\textbf{Z}' \boldsymbol{\beta_{\rm NT}}) \mathrm{exp}\{-\gamma_{\rm NT} t^{\alpha_{\rm NT}}
\mathrm{exp}(\textbf{Z}' \boldsymbol{\beta_{\rm NT}} )\}
$$
$$
\times \gamma_{\rm OS} \alpha_{\rm OSNT} y_{\rm OS}^{{\alpha_{\rm OSNT} -1}} \exp(\textbf{Z}' \boldsymbol{\beta_{\rm OSNT}}) \mathrm{exp}\{-\gamma_{\rm OS} t^{\alpha_{\rm OSNT}}
\mathrm{exp}(\textbf{Z}' \boldsymbol{\beta_{\rm OSNT}} )\}
$$

and

$$
L_2=-\frac{\partial S_1(t_{\rm NT}, t_{\rm OS}|\textbf{Z})}{\partial t_{\rm NT}}\bigg|_{t_{\rm NT}=y_{\rm NT},t_{\rm OS}=y_{\rm OS}}
$$
$$
=\bigg(\mathrm{exp}\{-\gamma_{\rm NT} y_{\rm NT}^{\alpha_{\rm NT}}
\mathrm{exp}(\textbf{Z}' \boldsymbol{\beta_{\rm NT}} )\}\bigg)^{-(\eta_{\rm NT}+1)}
$$
$$
\times \bigg(\mathrm{exp}\{-\gamma_{\rm NT} y_{\rm NT}^{\alpha_{\rm NT}}
\mathrm{exp}(\textbf{Z}' \boldsymbol{\beta_{\rm NT}} )\}^{-\eta_{\rm NT}}+\mathrm{exp}\{-\gamma_{\rm OS} y_{\rm OS}^{\alpha_{\rm OSNT}}
\mathrm{exp}(\textbf{Z}' \boldsymbol{\beta_{\rm OSNT}} )\}^{-\eta_{\rm NT}}-1\bigg)^{-\frac{\eta_{\rm NT}+1}{\eta_{\rm NT}}}
$$
$$
\times \gamma_{\rm NT} \alpha_{\rm NT} y_{\rm NT}^{{\alpha_{\rm NT} -1}} \exp(\textbf{Z}' \boldsymbol{\beta_{\rm NT}}) \mathrm{exp}\{-\gamma_{\rm NT} t^{\alpha_{\rm NT}}
\mathrm{exp}(\textbf{Z}' \boldsymbol{\beta_{\rm NT}} )\}
$$

and

$$
L_3=-\frac{\partial S_1(t_{\rm NT}, t_{\rm OS}|\textbf{Z})}{\partial t_{\rm OS}}\bigg|_{t_{\rm NT}=y_{\rm NT},t_{\rm OS}=y_{\rm OS}}
$$
$$
=\bigg(\mathrm{exp}\{-\gamma_{\rm OS} y_{\rm OS}^{\alpha_{\rm OSNT}}
\mathrm{exp}(\textbf{Z}' \boldsymbol{\beta_{\rm OSNT}} )\}\bigg)^{-(\eta_{\rm NT}+1)}
$$
$$
\times \bigg(\mathrm{exp}\{-\gamma_{\rm NT} y_{\rm NT}^{\alpha_{\rm NT}}
\mathrm{exp}(\textbf{Z}' \boldsymbol{\beta_{\rm NT}} )\}^{-\eta_{\rm NT}}+\mathrm{exp}\{-\gamma_{\rm OS} y_{\rm OS}^{\alpha_{\rm OSNT}}
\mathrm{exp}(\textbf{Z}' \boldsymbol{\beta_{\rm OSNT}} )\}^{-\eta_{\rm NT}}-1\bigg)^{-\frac{\eta_{\rm NT}+1}{\eta_{\rm NT}}}
$$
$$
\times \gamma_{\rm OS} \alpha_{\rm OSNT} y_{\rm OS}^{{\alpha_{\rm OSNT} -1}} \exp(\textbf{Z}' \boldsymbol{\beta_{\rm OSNT}}) \mathrm{exp}\{-\gamma_{\rm OS} t^{\alpha_{\rm OSNT}}
\mathrm{exp}(\textbf{Z}' \boldsymbol{\beta_{\rm OSNT}} )\}
$$

and

$$
L_4=S_1(t_{\rm NT}, t_{\rm OS}|\textbf{Z})\bigg|_{t_{\rm NT}=y_{\rm NT},t_{\rm OS}=y_{\rm OS}}
$$
$$
=\{\mathrm{exp}\{-\gamma_{\rm NT} y_{\rm NT}^{\alpha_{\rm NT}}
\mathrm{exp}(\textbf{Z}' \boldsymbol{\beta_{\rm NT}} )\}^{-\eta_{\rm NT}}
+\mathrm{exp}\{-\gamma_{\rm OS} y_{\rm OS}^{\alpha_{\rm OSNT}}
\mathrm{exp}(\textbf{Z}' \boldsymbol{\beta_{\rm OSNT}} )\}^{-\eta_{\rm NT}}-1\}^{-1/\eta_{\rm NT}}
$$
For a given subject, L1, L2, L3 and L4 correspond to the likelihood components that both NT and OS are observed, NT is observed but OS is censored, NT is censored but OS is observed, and both NT and OS are censored, respectively.

\newpage
\section{Likelihood derivation for the copula model of the new lesion and OS}
For new lesion, we define $y_{\rm NL}=\mathrm{min}(t_{\rm NL}, c_{\rm NL})$ and
$\delta_{\rm NL}=I(t_{\rm NL} \le c_{\rm NL})$,
where $c_{\rm NL}$ and $I(\cdot)$ denote the censoring time and the indicator function,
respectively; and define $y_{\rm OS}$ and $\delta_{\rm OS}$ similarly for OS.

Depending on the censoring pattern, the observed data for the $i$th participant falls into one of the following mutually exclusive cases: 1) both $t_{NL}$ and $t_{OS}$ are observed ($\delta_{NL}$=1, $\delta_{OS}$=1), 2) $t_{NL}$ is observed and $t_{OS}$ is censored ($\delta_{NL}$=1, $\delta_{OS}$=0), 3) $t_{NL}$ is censored and $t_{OS}$ is observed ($\delta_{NL}$=0, $\delta_{OS}$=1), and 4) both $t_{NL}$ and $t_{OS}$ are censored ($\delta_{NL}$=0, $\delta_{OS}$=0). Based on these four scenarios, we derive the likelihood of the copula model. 
Let
$\boldsymbol{\theta}_3=(\boldsymbol{\beta_{\rm NL}}, \alpha_{\rm NL}, \lambda_{\rm NL}, \boldsymbol{\beta_{\rm OSNL}},
\alpha_{\rm OSNL}, \lambda_{\rm OS}, \eta_{\rm NL})$,  then the likelihood for the $i$th patient with ${\rm Data}_i=(y_{\rm NL}, y_{\rm OS},
\delta_{\rm NL}, \delta_{\rm OS}, \textbf{Z})_i$ is given by
$$
L(\boldsymbol{\theta}_3|{\rm Data}_i)
=L_1^{\delta_{\rm NL}\delta_{\rm OS}}
L_2^{\delta_{\rm NL}(1-\delta_{\rm OS})}L_3^{(1-\delta_{\rm NL})\delta_{\rm OS}}
L_4^{(1-\delta_{\rm NL})(1-\delta_{\rm OS})}, 
$$
where
$$
L_1=\frac{\partial^2\, S_2(t_{\rm NL}, t_{\rm OS}|\textbf{Z})}{\partial t_{\rm NL} \partial t_{\rm OS}}\big|_{t_{\rm NL}=y_{\rm NL},t_{\rm OS}=y_{\rm OS}}
$$
$$
=\bigg(\eta_{\rm NL}+1\bigg)\bigg(\mathrm{exp}\{-\gamma_{\rm NL} y_{\rm NL}^{\alpha_{\rm NL}}
\mathrm{exp}(\textbf{Z}' \boldsymbol{\beta_{\rm NL}} )\} \mathrm{exp}\{-\gamma_{\rm OS} y_{\rm OS}^{\alpha_{\rm OSNL}}
\mathrm{exp}(\textbf{Z}' \boldsymbol{\beta_{\rm OSNL}} )\}\bigg)^{-(\eta_{\rm NL}+1)}
$$
$$
\times \bigg(\mathrm{exp}\{-\gamma_{\rm NL} y_{\rm NL}^{\alpha_{\rm NL}}
\mathrm{exp}(\textbf{Z}' \boldsymbol{\beta_{\rm NL}} )\}^{-\eta_{\rm NL}}+\mathrm{exp}\{-\gamma_{\rm OS} y_{\rm OS}^{\alpha_{\rm OSNL}}
\mathrm{exp}(\textbf{Z}' \boldsymbol{\beta_{\rm OSNL}} )\}^{-\eta_{\rm NL}}-1\bigg)^{-\frac{2\eta_{\rm NL}+1}{\eta_{\rm NL}}}
$$
$$
\times \gamma_{\rm NL} \alpha_{\rm NL} y_{\rm NL}^{{\alpha_{\rm NL} -1}} \exp(\textbf{Z}' \boldsymbol{\beta_{\rm NL}}) \mathrm{exp}\{-\gamma_{\rm NL} t^{\alpha_{\rm NL}}
\mathrm{exp}(\textbf{Z}' \boldsymbol{\beta_{\rm NL}} )\}
$$
$$
\times \gamma_{\rm OS} \alpha_{\rm OSNL} y_{\rm OS}^{{\alpha_{\rm OSNL} -1}} \exp(\textbf{Z}' \boldsymbol{\beta_{\rm OSNL}}) \mathrm{exp}\{-\gamma_{\rm OS} t^{\alpha_{\rm OSNL}}
\mathrm{exp}(\textbf{Z}' \boldsymbol{\beta_{\rm OSNL}} )\}
$$

and

$$
L_2=-\frac{\partial S_2(t_{\rm NL}, t_{\rm OS}|\textbf{Z})}{\partial t_{\rm NL}}\bigg|_{t_{\rm NL}=y_{\rm NL},t_{\rm OS}=y_{\rm OS}}
$$
$$
=\bigg(\mathrm{exp}\{-\gamma_{\rm NL} y_{\rm NL}^{\alpha_{\rm NL}}
\mathrm{exp}(\textbf{Z}' \boldsymbol{\beta_{\rm NL}} )\}\bigg)^{-(\eta_{\rm NL}+1)}
$$
$$
\times \bigg(\mathrm{exp}\{-\gamma_{\rm NL} y_{\rm NL}^{\alpha_{\rm NL}}
\mathrm{exp}(\textbf{Z}' \boldsymbol{\beta_{\rm NL}} )\}^{-\eta_{\rm NL}}+\mathrm{exp}\{-\gamma_{\rm OS} y_{\rm OS}^{\alpha_{\rm OSNL}}
\mathrm{exp}(\textbf{Z}' \boldsymbol{\beta_{\rm OSNL}} )\}^{-\eta_{\rm NL}}-1\bigg)^{-\frac{\eta_{\rm NL}+1}{\eta_{\rm NL}}}
$$
$$
\times \gamma_{\rm NL} \alpha_{\rm NL} y_{\rm NL}^{{\alpha_{\rm NL} -1}} \exp(\textbf{Z}' \boldsymbol{\beta_{\rm NL}}) \mathrm{exp}\{-\gamma_{\rm NL} t^{\alpha_{\rm NL}}
\mathrm{exp}(\textbf{Z}' \boldsymbol{\beta_{\rm NL}} )\}
$$

and

$$
L_3=-\frac{\partial S_2(t_{\rm NL}, t_{\rm OS}|\textbf{Z})}{\partial t_{\rm OS}}\bigg|_{t_{\rm NL}=y_{\rm NL},t_{\rm OS}=y_{\rm OS}}
$$
$$
=\bigg(\mathrm{exp}\{-\gamma_{\rm OS} y_{\rm OS}^{\alpha_{\rm OSNL}}
\mathrm{exp}(\textbf{Z}' \boldsymbol{\beta_{\rm OSNL}} )\}\bigg)^{-(\eta_{\rm NL}+1)}
$$
$$
\times \bigg(\mathrm{exp}\{-\gamma_{\rm NL} y_{\rm NL}^{\alpha_{\rm NL}}
\mathrm{exp}(\textbf{Z}' \boldsymbol{\beta_{\rm NL}} )\}^{-\eta_{\rm NL}}+\mathrm{exp}\{-\gamma_{\rm OS} y_{\rm OS}^{\alpha_{\rm OSNL}}
\mathrm{exp}(\textbf{Z}' \boldsymbol{\beta_{\rm OSNL}} )\}^{-\eta_{\rm NL}}-1\bigg)^{-\frac{\eta_{\rm NL}+1}{\eta_{\rm NL}}}
$$
$$
\times \gamma_{\rm OS} \alpha_{\rm OSNL} y_{\rm OS}^{{\alpha_{\rm OSNL} -1}} \exp(\textbf{Z}' \boldsymbol{\beta_{\rm OSNL}}) \mathrm{exp}\{-\gamma_{\rm OS} t^{\alpha_{\rm OSNL}}
\mathrm{exp}(\textbf{Z}' \boldsymbol{\beta_{\rm OSNL}} )\}
$$

and

$$
L_4=S_1(t_{\rm NL}, t_{\rm OS}|\textbf{Z})\bigg|_{t_{\rm NL}=y_{\rm NL},t_{\rm OS}=y_{\rm OS}}
$$
$$
=\{\mathrm{exp}\{-\gamma_{\rm NL} y_{\rm NL}^{\alpha_{\rm NL}}
\mathrm{exp}(\textbf{Z}' \boldsymbol{\beta_{\rm NL}} )\}^{-\eta_{\rm NL}}
+\mathrm{exp}\{-\gamma_{\rm OS} y_{\rm OS}^{\alpha_{\rm OSNL}}
\mathrm{exp}(\textbf{Z}' \boldsymbol{\beta_{\rm OSNL}} )\}^{-\eta_{\rm NL}}-1\}^{-1/\eta_{\rm NL}}
$$
For a given subject, L1, L2, L3 and L4 correspond to the likelihood components that both NL and OS are observed, NL is observed but OS is censored, NL is censored but OS is observed, and both NL and OS are censored, respectively.

\newpage
\section{Likelihood derivation for the copula model between TTP and OS}
For TTP,
define $y_{\rm TTP}=\mathrm{min}(t_{\rm TTP}, c_{\rm TTP})$ and
$\delta_{\rm TTP}=I(t_{\rm TTP} \le c_{\rm TTP})$,
where $c_{\rm TTP}$ and $I(\cdot)$ denote the censoring time and the indicator function,
respectively; and define $y_{\rm OS}$ and $\delta_{\rm OS}$ similarly for OS. Depending on the censoring pattern, the observed data for the $i$th participant falls into one of the four mutually exclusive cases: ($\delta_{TTP}$=1, $\delta_{OS}$=1), ($\delta_{TTP}$=1, $\delta_{OS}$=0),
($\delta_{TTP}$=0, $\delta_{OS}$=1), and ($\delta_{TTP}$=0, $\delta_{OS}$=0). 

Let
$\boldsymbol{\theta}_{TTP}=(\boldsymbol{\beta_{\rm TTP}}, \alpha_{\rm TTP}, \lambda_{\rm TTP}, \boldsymbol{\beta_{\rm OSTTP}},
\alpha_{\rm OSTTP}, \lambda_{\rm OS}, \eta_{\rm TTP})$,  then the
likelihood for the $i$th patient with ${\rm Data}_i=(y_{\rm TTP}, y_{\rm OS},
\delta_{\rm TTP}, \delta_{\rm OS}, \textbf{Z})_i$ is given by
$$
L(\boldsymbol{\theta}_{TTP}|{\rm Data}_i)
=L_1^{\delta_{\rm TTP}\delta_{\rm OS}}
L_2^{\delta_{\rm TTP}(1-\delta_{\rm OS})}L_3^{(1-\delta_{\rm TTP})\delta_{\rm OS}}
L_4^{(1-\delta_{\rm TTP})(1-\delta_{\rm OS})},
$$

where
$$
L_1=\frac{\partial^2\, S_2(t_{\rm TTP}, t_{\rm OS}|\textbf{Z})}{\partial t_{\rm TTP} \partial t_{\rm OS}}\big|_{t_{\rm TTP}=y_{\rm TTP},t_{\rm OS}=y_{\rm OS}}
$$
$$
=\bigg(\eta_{\rm TTP}+1\bigg)\bigg(\mathrm{exp}\{-\gamma_{\rm TTP} y_{\rm TTP}^{\alpha_{\rm TTP}}
\mathrm{exp}(\textbf{Z}' \boldsymbol{\beta_{\rm TTP}} )\} \mathrm{exp}\{-\gamma_{\rm OS} y_{\rm OS}^{\alpha_{\rm OSTTP}}
\mathrm{exp}(\textbf{Z}' \boldsymbol{\beta_{\rm OSTTP}} )\}\bigg)^{-(\eta_{\rm TTP}+1)}
$$
$$
\times \bigg(\mathrm{exp}\{-\gamma_{\rm TTP} y_{\rm TTP}^{\alpha_{\rm TTP}}
\mathrm{exp}(\textbf{Z}' \boldsymbol{\beta_{\rm TTP}} )\}^{-\eta_{\rm TTP}}+\mathrm{exp}\{-\gamma_{\rm OS} y_{\rm OS}^{\alpha_{\rm OSTTP}}
\mathrm{exp}(\textbf{Z}' \boldsymbol{\beta_{\rm OSTTP}} )\}^{-\eta_{\rm TTP}}-1\bigg)^{-\frac{2\eta_{\rm TTP}+1}{\eta_{\rm TTP}}}
$$
$$
\times \gamma_{\rm TTP} \alpha_{\rm TTP} y_{\rm TTP}^{{\alpha_{\rm TTP} -1}} \exp(\textbf{Z}' \boldsymbol{\beta_{\rm TTP}}) \mathrm{exp}\{-\gamma_{\rm TTP} t^{\alpha_{\rm TTP}}
\mathrm{exp}(\textbf{Z}' \boldsymbol{\beta_{\rm TTP}} )\}
$$
$$
\times \gamma_{\rm OS} \alpha_{\rm OSTTP} y_{\rm OS}^{{\alpha_{\rm OSTTP} -1}} \exp(\textbf{Z}' \boldsymbol{\beta_{\rm OSTTP}}) \mathrm{exp}\{-\gamma_{\rm OS} t^{\alpha_{\rm OSTTP}}
\mathrm{exp}(\textbf{Z}' \boldsymbol{\beta_{\rm OSTTP}} )\}
$$

and

$$
L_2=-\frac{\partial S_2(t_{\rm TTP}, t_{\rm OS}|\textbf{Z})}{\partial t_{\rm TTP}}\bigg|_{t_{\rm TTP}=y_{\rm TTP},t_{\rm OS}=y_{\rm OS}}
$$
$$
=\bigg(\mathrm{exp}\{-\gamma_{\rm TTP} y_{\rm TTP}^{\alpha_{\rm TTP}}
\mathrm{exp}(\textbf{Z}' \boldsymbol{\beta_{\rm TTP}} )\}\bigg)^{-(\eta_{\rm TTP}+1)}
$$
$$
\times \bigg(\mathrm{exp}\{-\gamma_{\rm TTP} y_{\rm TTP}^{\alpha_{\rm TTP}}
\mathrm{exp}(\textbf{Z}' \boldsymbol{\beta_{\rm TTP}} )\}^{-\eta_{\rm TTP}}+\mathrm{exp}\{-\gamma_{\rm OS} y_{\rm OS}^{\alpha_{\rm OSTTP}}
\mathrm{exp}(\textbf{Z}' \boldsymbol{\beta_{\rm OSTTP}} )\}^{-\eta_{\rm TTP}}-1\bigg)^{-\frac{\eta_{\rm TTP}+1}{\eta_{\rm TTP}}}
$$
$$
\times \gamma_{\rm TTP} \alpha_{\rm TTP} y_{\rm TTP}^{{\alpha_{\rm TTP} -1}} \exp(\textbf{Z}' \boldsymbol{\beta_{\rm TTP}}) \mathrm{exp}\{-\gamma_{\rm TTP} t^{\alpha_{\rm TTP}}
\mathrm{exp}(\textbf{Z}' \boldsymbol{\beta_{\rm TTP}} )\}
$$

and

$$
L_3=-\frac{\partial S_2(t_{\rm TTP}, t_{\rm OS}|\textbf{Z})}{\partial t_{\rm OS}}\bigg|_{t_{\rm TTP}=y_{\rm TTP},t_{\rm OS}=y_{\rm OS}}
$$
$$
=\bigg(\mathrm{exp}\{-\gamma_{\rm OS} y_{\rm OS}^{\alpha_{\rm OSTTP}}
\mathrm{exp}(\textbf{Z}' \boldsymbol{\beta_{\rm OSTTP}} )\}\bigg)^{-(\eta_{\rm TTP}+1)}
$$
$$
\times \bigg(\mathrm{exp}\{-\gamma_{\rm TTP} y_{\rm TTP}^{\alpha_{\rm TTP}}
\mathrm{exp}(\textbf{Z}' \boldsymbol{\beta_{\rm TTP}} )\}^{-\eta_{\rm TTP}}+\mathrm{exp}\{-\gamma_{\rm OS} y_{\rm OS}^{\alpha_{\rm OSTTP}}
\mathrm{exp}(\textbf{Z}' \boldsymbol{\beta_{\rm OSTTP}} )\}^{-\eta_{\rm TTP}}-1\bigg)^{-\frac{\eta_{\rm TTP}+1}{\eta_{\rm TTP}}}
$$
$$
\times \gamma_{\rm OS} \alpha_{\rm OSTTP} y_{\rm OS}^{{\alpha_{\rm OSTTP} -1}} \exp(\textbf{Z}' \boldsymbol{\beta_{\rm OSTTP}}) \mathrm{exp}\{-\gamma_{\rm OS} t^{\alpha_{\rm OSTTP}}
\mathrm{exp}(\textbf{Z}' \boldsymbol{\beta_{\rm OSTTP}} )\}
$$

and

$$
L_4=S_1(t_{\rm TTP}, t_{\rm OS}|\textbf{Z})\bigg|_{t_{\rm TTP}=y_{\rm TTP},t_{\rm OS}=y_{\rm OS}}
$$
$$
=\{\mathrm{exp}\{-\gamma_{\rm TTP} y_{\rm TTP}^{\alpha_{\rm TTP}}
\mathrm{exp}(\textbf{Z}' \boldsymbol{\beta_{\rm TTP}} )\}^{-\eta_{\rm TTP}}
+\mathrm{exp}\{-\gamma_{\rm OS} y_{\rm OS}^{\alpha_{\rm OSTTP}}
\mathrm{exp}(\textbf{Z}' \boldsymbol{\beta_{\rm OSTTP}} )\}^{-\eta_{\rm TTP}}-1\}^{-1/\eta_{\rm TTP}}
$$
For a given subject, $L_1, L_2$,  $L_3$ and $L_4$ correspond to the likelihood components that both TTP and OS are observed, TTP is observed but OS is censored, TTP is censored but OS is observed, and both TTP and OS are censored, respectively.

\newpage
\begin{figure}[H]
    \centering
    \includegraphics[width=0.8\textwidth]{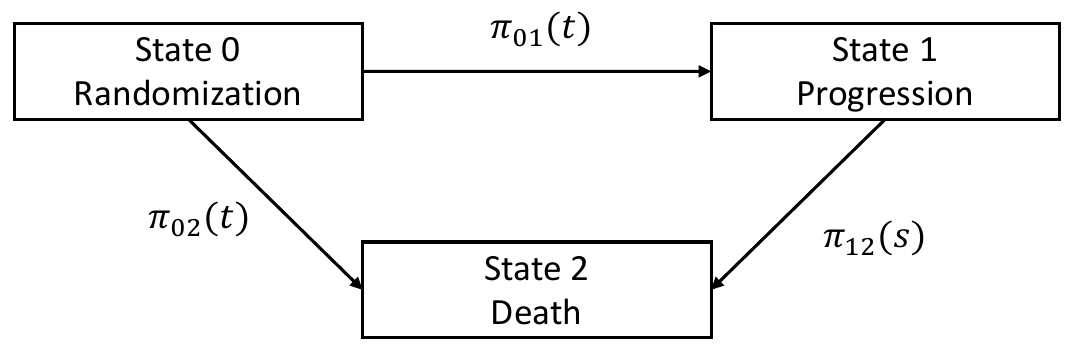}
    \caption{The multi-state (three-state illness-death) model \label{fig:multistate}}
\end{figure}

\newpage
\section{Likelihood derivation for the multi-state model}
For the multi-state model, we aim to estimate the parameter vector
\[
\theta = (\alpha, \gamma_{01}, \gamma_{02}, \gamma_{12})
\]
using maximum likelihood estimation.

\textbf{Modeling the Survival Experiences}:

Individual survival experiences can be characterized by four distinct cases based on a patient's progression through the multi-state model:

\begin{enumerate}
    \item Patients progress and are then censored.
    \item Patients progress and subsequently die.
    \item Patients die without prior progression.
    \item Patients are censored without experiencing either progression or death.
\end{enumerate}

For every individual \( i \) in the set \( (1, ..., n) \):

\begin{itemize}
    \item \( t_{i1} \) indicates the duration from the initial state to either progression or death.
    \item \( t_{i2} \), which is conditioned on progression, represents the time from progression to death.
\end{itemize}

\textbf{Likelihood Estimation for Individual Experiences}:

The individual likelihood for these experiences, denoted as \( L_i^{(k)}(\theta) \) where \( k \) varies from 1 to 4, can be described as:

\[
L_i^{(1)}(\theta) = f_1(t_{i1})S_2(t_{i1})S_3(t_{i2})
\]

\[
L_i^{(2)}(\theta) = f_1(t_{i1})S_2(t_{i1})f_3(t_{i2})
\]

\[
L_i^{(3)}(\theta) = S_1(t_{i1})f_2(t_{i1})
\]

\[
L_i^{(4)}(\theta) = S_1(t_{i1})S_2(t_{i1})
\]

Here:

\begin{itemize}
    \item \( f_1(\cdot) \) and \( S_1(\cdot) \): Density and survival function for time to progression.
    \item \( f_2(\cdot) \) and \( S_2(\cdot) \): Density and survival function for time to death without progression.
    \item \( f_3(\cdot) \) and \( S_3(\cdot) \): Density and survival function for time to death post progression.
\end{itemize}

\textbf{Overall Log-Likelihood}:

The comprehensive log-likelihood for all subjects can be formulated as:

\[
\begin{aligned}
\log(L(\theta)) &= \sum_{i=1}^{n}(d_1)(1-d_2)(1-d_3)L_i^{(1)}(\theta) \\
&\quad + (d_1)(1-d_2)(d_3)L_i^{(2)}(\theta) \\
&\quad +(1-d_1)(d_2)L_i^{(3)}(\theta) \\
&\quad +(1-d_1)(1-d_2)L_i^{(4)}(\theta),
\end{aligned}
\]

Where:

\begin{itemize}
    \item \( d_1 \): Indicator for progression.
    \item \( d_2 \): Indicator for death without preceding progression.
    \item \( d_3 \): Indicator for death post progression.
\end{itemize}

Note: An indicator value of 0 represents censoring, whereas a value of 1 indicates the event's occurrence.

\begin{table}
\caption{The coverage rate (CR) and credible interval (CI) width for predictors of the time of the last death. Comparing the results from the multivariate joint modeling approach (baveJM), a shared-parameter random-effect joint model (SPJM), a copula model between TTP and OS (Cop), , a multi-state model (MS), and the marginal Weibull baseline hazard model of OS (Mrgl). The unit for CI width is month.}
\label{tab:sim_result}
\begin{center}
\begin{tabular}{ccccccccccc}
& \multicolumn{5}{c}{Independent OS} & \multicolumn{5}{c}{OS dep. on NL} \\ 
& baveJM & SPJM & Cop & Mrgl & MS & baveJM & SPJM & Cop & Mrgl & MS \\ \hline
\multicolumn{11}{l}{\textit{100 death event}}\\
CR & 0.78 & 1.00 & 0.86 & 0.00 & 0.83 & 0.30 & 1.00 & 0.85 & 0.00 & 0.93 \\ 
Width & 53.2 & 68.1 & 54.7 & 17.1 & 55.4 & 38.2 & 68.6 & 54.6 & 16.9 & 59.7 \\ \hline
\multicolumn{11}{l}{\textit{200 death event}}\\
CR & 0.96 & 0.89 & 0.97 & 0.00 & 0.61 & 0.92 & 0.89 & 0.98 & 0.00 & 0.78 \\ 
Width & 54.0 & 50.9 & 54.7 & 20.0 & 42.4 & 54.6 & 51.8 & 54.6 & 20.0 & 46.1 \\ \hline
\multicolumn{11}{l}{\textit{300 death event}}\\
CR & 1.00 & 0.70 & 1.00 & 0.01 & 0.57 & 0.96 & 0.71 & 0.99 & 0.00 & 0.74 \\ 
Width & 49.8 & 37.6 & 49.9 & 19.8 & 34.3 & 50.0 & 38.4 & 49.8 & 19.4 & 37.5  \\ 
\hline
\multicolumn{11}{l}{\textit{350 death event}}\\
CR & 1.00 & 0.68 & 1.00 & 0.06 & 0.65 & 0.95 & 0.66 & 1.00 & 0.02 & 0.76 \\ 
Width & 45.3 & 31.4 & 45.6 & 18.4 & 29.7 & 42.1 & 31.9 & 45.5 & 18.1 & 32.8 \\ 
\hline
\end{tabular}
\end{center}
\end{table}

\begin{table}
\begin{center}
\begin{tabular}{ccccccccccc}
& \multicolumn{5}{c}{OS dep. on TL} & \multicolumn{5}{c}{OS dep. on both NL and TL} \\ 
& baveJM & SPJM & Cop & Mrgl & MS & baveJM & SPJM & Cop & Mrgl & MS \\ \hline
\multicolumn{11}{l}{\textit{100 death event}}\\
CR & 0.96 & 1.00 & 0.05 & 0.00 & 0.40 & 0.91 & 1.00 & 0.92 & 0.00 & 1.00 \\ 
Width & 73.2 & 85.0 & 43.9 & 9.8 & 47.7 & 64.6 & 71.6 & 55.0 & 18.8 & 66.6  \\ \hline
\multicolumn{11}{l}{\textit{200 death event}}\\
CR & 0.83 & 1.00 & 0.00 & 0.00 & 0.00 & 0.99 & 1.00 & 0.87 & 0.00 & 0.98 \\ 
Width & 58.0 & 74.9 & 30.1 & 8.2 & 20.6 & 58.7 & 68.9 & 48.4 & 16.3  & 53.7 \\ 
\hline
\multicolumn{11}{l}{\textit{300 death event}}\\
CR & 0.97 & 1.00 & 0.04 & 0.00 & 0.00 & 0.98 & 1.00 & 0.80 & 0.00 & 0.90 \\ 
Width & 59.3 & 65.0 & 34.7 & 12.6 & 18.6 & 47.5 & 59.7 & 38.9 & 15.0 & 40.1  \\ 
\hline
\multicolumn{11}{l}{\textit{350 death event}}\\
CR & 1.00 & 1.00 & 0.77 & 0.00 & 0.00 & 0.96 & 1.00 & 0.87 & 0.00 & 0.89 \\ 
Width & 55.9 & 59.6 & 41.7 & 16.8 & 20.2 & 38.9 & 60.0 & 34.6 & 14.2 & 34.2  \\ 
\hline
\end{tabular}
\end{center}
\end{table}

\newpage
\begin{figure}[H]
\centering
\includegraphics[width=0.7\textwidth]{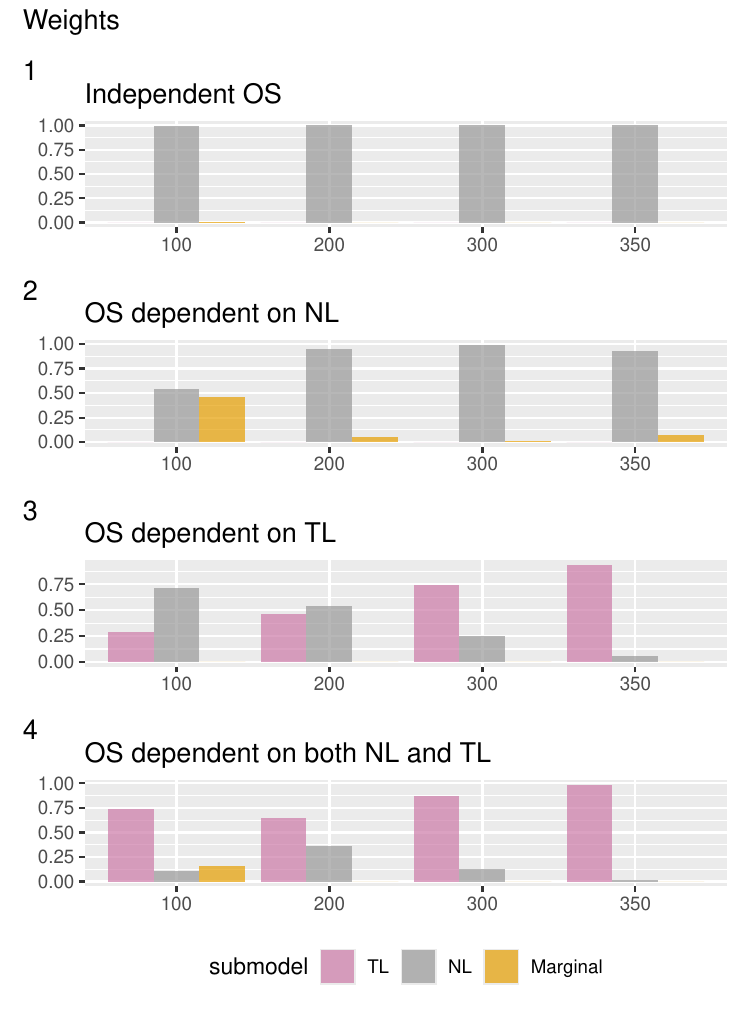}
\caption{The weights of all submodels of the multivariate joint modeling approach for each snapshot dataset in the simulation studies.\label{fig:weights_simulation}}
\end{figure}

\newpage









